\documentclass[fleqn,usenatbib]{mnras}

\usepackage{newtxtext,newtxmath}

\usepackage[T1]{fontenc}
\usepackage{ae,aecompl}

\usepackage{graphicx}	
\usepackage{subfigure}
\usepackage{bm}
\usepackage{gensymb}
\usepackage{enumitem}
\usepackage{xspace}
\usepackage{multirow}
\usepackage[dvipsnames]{xcolor}
\usepackage[utf8]{inputenc}
\usepackage{lscape}

\newcommand{\msun}{M$_\odot$}

\newcommand{\chmk}{\checkmark}

\title[An old cluster in Eridanus II]{Star cluster survival in dark matter halos: An old cluster in Eridanus II?}

\author[J. Alzate, V. Lora, G. Bruzual, L. Lomel\'\i-N\'u\~nez \& B. Cervantes Sodi]{
Jairo A. Alzate,$^{1}$\thanks{E-mail: j.alzate@irya.unam.mx}
Ver\'onica Lora,$^{1}$
Gustavo Bruzual,$^{1}$
Luis Lomel\'i-N\'u\~nez,$^{2}$
\newauthor           
Bernardo Cervantes Sodi$^{1}$ 
\\
$^{1}$Instituto de Radioastronom{\'i}a y Astrof{\'i}sica, UNAM, Campus Morelia, Michoac\'an, C.P. 58089, M{\'e}xico\\
$^{2}$Instituto Nacional de Astrof{\'i}sica, Optica y Electr\'onica, Calle Luis Enrique Erro, No 1, Tonantzintla, Puebla, C.P. 72840, M\'exico
}

\date{Accepted XXX. Received YYY; in original form ZZZ}

\pubyear{2020}

\begin{document}
\label{firstpage}
\pagerange{\pageref{firstpage}--\pageref{lastpage}}
\maketitle

\begin{abstract}
The star formation history and the internal dynamics of Milky Way satellite galaxies are often complicated.
In the last years, a substantial fraction of the known faint dwarf satellites have been studied.
Some of them show embedded stellar substructures, such as star clusters and even globular star clusters.
In this work we study Eridanus II, a dwarf spheroidal satellite which hosts a star cluster, using published and archival data 
from the Hubble Space Telescope Advanced Camera for Surveys.
We employ a Bayesian hierarchical method to infer the star formation history of Eridanus II.
We find that the bulk of the stars in Eridanus II are very old ($13.5_{-1}^{+0.5}$ Gyr) and quite metal poor ($Z$\,=\,$0.00001$).
We do not find any evidence of the presence of an intermediate age or young population in Eri II.
We cannot date the embedded star cluster as a separate entity, but we find it likely that the cluster has a similar age and metallicity as the bulk of the stars in Eri II.

The existence of an old star cluster in a dark matter dominated old metal poor dwarf galaxy 
is of major importance to cast light on the dark matter distribution within dwarf galaxies. 
The existence of intermediate age stars is required by the recent detection of carbon stars in Eri II.
Since no recent star formation is detected, {\it blue-straggler} fusions of lower mass stars are the most likely origin of the carbon star progenitors.

\end{abstract}



\section{Introduction} 

\begin{figure*}
\begin{center}
    \includegraphics[width=1.8\columnwidth]{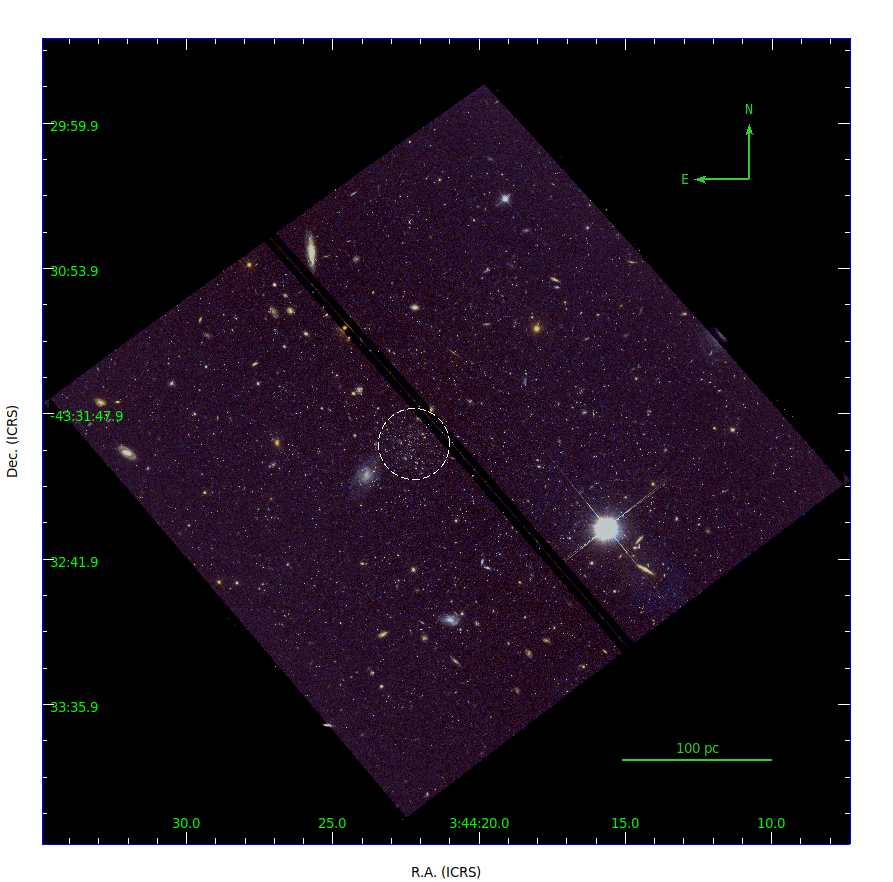}
    \caption{\label{fig:eri_img}
    HST/ACS image of Eri II in the WFC F814W band. The {\it dashed} circle encloses the structure identidied as a star cluster.}
\end{center}
\end{figure*}

Dwarf galaxies are the most numerous galaxies in the Universe. They come in different flavors, dwarf spheroidals (dSphs) being just one of them. Because dSphs lack gas and most of their stars are old \citep[age $>\,10$ Gyr,][]{grebel:04}, they were once considered simple stellar populations (SSPs). As better observations have been obtained and more research accomplished, it has become clear that these galaxies contain more than meets the eye. To study dwarf galaxies it is a natural choice to start with the satellite galaxies of the Milky Way (MW), which by their proximity can be resolved into individual stars and their evolutionary history traced in great detail.
\cite{munoz:18} carried out a wide-field imaging survey of the MW satellites in the outer halo (Galactocentric distance $>25$ kpc) and derived structural parameters for 58 galaxies, 75\% of the known outer halo satellites. 

One particularly interesting MW satellite is the dSph galaxy Eridanus II \citep[Eri II hereafter, shown in Fig.\,\ref{fig:eri_img};][]{bechtol:15, koposov:15, crnojevic:16, Li:17, zout:20,simon:21,gallart:21}\footnote{The papers by \cite{simon:21} and \cite{gallart:21} were submitted after this paper and their content was unknown to us in advance of publication.}.
At a Galactocentric distance of $368 \pm 17$ kpc \citep[distance modulus\,=\,22.83][]{crnojevic:16}, 
Eri II is beyond the $300$ kpc virial radius of the MW. 
Using \textit{Magellan}/IMACS spectroscopy, \cite{Li:17} measured a velocity dispersion of 6.9$^{+1.2}_{-0.9}$ km s$^{-1}$ for Eri II.
For a velocity dispersion supported system in equilibrium, this implies a mass within the half-light radius of 1.2$^{+0.4}_{-0.3} \times 10^7$M$_{\odot}$, and a mass-to-light ratio of $420$\ M$_{\odot}/L_{\odot}$, 
making Eri II {\it a dark matter} (DM) {\it dominated satellite galaxy}.

Not many satellite galaxies are known to host globular clusters (GCs).
Only twelve out of $>76$ in the Local Group, and only three MW dSphs, Sagittarius, Fornax and Eri II host GCs \citep{grebel:16}. 
A small number of MW dSph satellites contain stellar substructures, e.g.,
Ursa Minor \citep{kleyna:98,kleyna:04,walker:06}, 
Sextans; \citep{battaglia:11,kim:19,lora:19}, and Carina; \citep{lora:19}.
Eri II is thus {\it one of the} MW dSph {\it galaxies that contains a star cluster}.

From Dark Energy Survey (DES) public data, \cite{koposov:15} identified a fuzzy object, a few parsecs in size, near the center of Eri II that they interpret as a GC. \cite{crnojevic:16} confirmed the presence of the partially resolved cluster. With M$_V=-3.5$, the cluster accounts for $4\%$ of Eri II's total luminosity (M$_V=-7.1$). Eri II is then {\it the least luminous galaxy known to host a star cluster}.

From spectroscopic observations with the Multi Unit Spectroscopic Explorer (MUSE) on the Very Large Telescope (VLT), \cite{zout:20} measured then mean line-of-sight velocity of 26 stars members of Eri II, 
seven of which were identified as possible cluster members.
From the velocity and velocity dispersion distributions they {\it confirm the existence} of the cluster, conclude that its population is dynamically separated from the bulk of the stars of Eri II, and find no evidence of an excess of dark matter associated to the cluster.

Evidence for the presence of several stellar populations in Eri II has been reported in the literature.
A major, old ($>10$ Gyr) stellar population component was identified in Eri II by \cite{crnojevic:16}, but it has been suggested that a younger population may also be present. 
\cite{koposov:15} argue in favor of a second population as young as 250 Myr, but this claim was refused by \cite{westmeirer:15}, who did not find any HI gas associated to Eri II. 
\cite{crnojevic:16} derived an HI mass limit of $M_{HI} < 2800$\ M$_{\odot}$, making Eri II {\it a very gas-poor galaxy.}
\cite{crnojevic:16} also report a possible $3$ Gyr old intermediate age population.
\cite{zout:20} assume an age of $8$ Gyr for both the bulk of the stars and the cluster in Eri II.
However, they find one confirmed and two candidates carbon (C) stars which indicate the presence of an intermediate-age population.

Analyzing deep HST phototometry \citet[][S21 hereafter]{simon:21} determined the SFH of Eri II and the structural parameters of the star cluster.
They show that at least 80\% of the stars of Eri II formed before $z$\,$\sim$\,$6$ (approximately 700 million years after the Big Bang).
From a statistical analysis of the sub-haloes in the ELVIS simulation \citep{garrison:14} they conclude that probably Eri II has not yet passed through its closest 
approach to the MW, and therefore that reionization is then the most likely cause for the quenching of star formation in Eri II.
However, \citet[][G21 hereafter]{gallart:21} consider more plausible that the quenching of star formation in Eri II is due to stellar feedback by supernova events than to reionization.
In favour of this idea they argue that the galaxy Leo T, with similar properties to Eri II, continued forming stars during a long period of time without being affected by reionization.
Both S21 and G21 derive the SFH of Eri II comparing synthetic colour-magnitude diagrams (CMDs) with data using histograms.
S21 use the statistical analysis of \citep{dolphin02} and G21 use the code TheStorm \citep{bernard15,bernard18}.

The existence of stellar substructure in dSph galaxies is of major importance. 
\cite{lora:12, lora:13} studied the survival of old kinematic stellar substructures (e.g., star clusters) embedded in the DM halo of the Fornax and Sextans dSph galaxies using N-body simulations.
The survival of stellar substructures within their DM halo suggests that the DM follows a core (flat) rather than a cuspy (NFW) central density profile \citep{nfw:97, lora:12, lora:13, amorisco:17, contenta:18}. 
The fact that Eri II is DM dominated and that it hosts a star cluster, makes it an ideal system to characterize its DM distribution.

In this paper we study in detail the stellar population(s) present in Eri II.
The variety of age estimates for the stars in this galaxy motivates our study.
To characterize the SFH of Eri II we avoid the isochrone fitting to the CMD and other data binning techniques, especially unappealing for sparse populations,
that have been used frequently in the literature \citep{ramirez:19}.
We opt for Bayesian inference, which allows us to estimate not only the physical parameters describing a stellar population but also their uncertainties.
We apply the Bayesian hierarchical method developed by \citet[A21 hereafter]{alzate21} to infer the age and metallicity distribution of resolved stellar populations,
to characterize the physical properties of the bulk of the stars present today in Eri II.
We attempt to date the star cluster in Eri II as a separate entity. Our method should detect intermediate age populations if present in Eri II.

The paper is organized as follows. 
In Section\,\ref{sec:data_reduction} we describe the HST data used in our analysis.
In Section\,\ref{sec:method} we summarize the Bayesian hierarchical statistical method developed elsewhere by us to infer the star formation history (SFH) of resolved stellar populations.
In Section\,\ref{sec:age_metal} we infer the age and metallicity distribution of the stars observed in Eri II.
In Appendix\,\ref{sec:appendix} we explore the ability of our method to detect or not more than one population of different age and metalliicty in a resolved galaxy and under what
conditions.
The conclusions are presented in Section\,\ref{sec:summary}.

\section{Eri II data} \label{sec:data_reduction}

In this work we use HST/ACS photometric data of Eri II in {\it (a)} the F606W and F814W bands published by S21, and {\it (b)} the F475W and F814W bands
derived by us from Hubble Legacy Archive images.

\subsection{F606W/F814W published photometry}

S21 include a photometric catalogue based on 20,680 s (F606W band) and 12,830 s (F814W band) exposures accumulated over 7 visits of the HST/ACS to Eri II as part of program GO-14234 (PI: J. D. Simon).
These long exposures resulted in the most accurate photometry of Eri II available to date, with a signal-to-noise ratio of 10 for stars 1 mag below the main sequence turn-off (MSTO), almost four times the signal-to-noise ratio achieved by G21. 
The S21 catalogue was derived using aperture and point spread function (PSF) photometry with the DAOPHOT-II package \citep{stetson:87} in the 
STmag\footnote{An object with constant flux $F_\lambda$\,=\,3.63\,$\times$,10$^{-9}$\,erg\,cm$^{-2}$\,s$^{-1}$\,\AA$^{-1}$ has magnitude STmag = 0 in every filter. STmag\,=\,-\,2.5\,log\,$F_\lambda$\,-\,21.1.} magnitude system.
The S21 catalogue is shown as a CMD in Fig.\,\ref{fig:eri_CMDs}b.

\subsection{F475W/F814W Hubble Legacy Archive images}

As part of program P.ID. 14224 \citep[Cycle 23, P.I. C. Gallart,][G16 hereafter]{gallart16}, deep photometric data were obtained during 2 visits of the HST/ACS to Eri II,
accumulating exposure times of $7644$ s (F475W band) and $7900$ s (F814W band). This proposal was aimed to build a good quality CMD of Eri II reaching the MSTO with enough precision and accuracy to unveil the SFH of this galaxy. The exposures were distributed such that the telescope could collect the light of the brighter stars without losing the fainter stars, avoiding CCD saturation and allowing to build a complete CMD for both the bright and faint regimes.
The long wavelength baseline provided by the filter pair (F475W, F814W) is ideal to study variations in age and metallicity in a resolved stellar population, requiring lower exposure times than other filter combinations to reach the necessary accuracy to separate the different stellar groups \citep[see, e.g.,][]{stetson:94}.

\begin{figure}
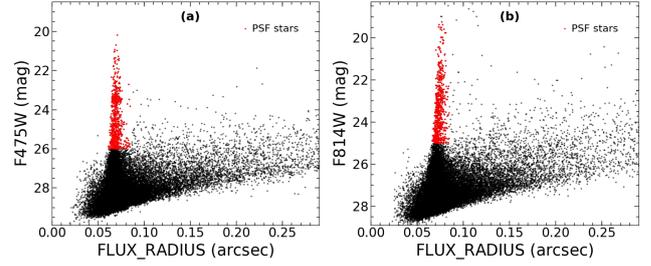

\begin{center}
        \includegraphics[width=0.49\columnwidth]{FIGURES/Fig_02_a_alzate_et_al_eridanus.pdf}
        \includegraphics[width=0.49\columnwidth]{FIGURES/Fig_02_b_alzate_et_al_eridanus.pdf}
        \caption{\label{fig:Init_Select}
        (\texttt{MAG\_AUTO}, \texttt{FLUX\_RADIUS}) plane for {\it (a)} F475W and {\it (b)} F814W.
        The {\it black points} represent the full initial catalogue and the {\it red points} the sources selected for the PSF model construction.}
\end{center}
\end{figure}

\begin{table}
\begin{center}
\caption{\label{tab:param_sel}Selection function for the PSF model.}
\begin{tabular}{lcccc}
\hline
\multirow{2}{*}{Filter}     & \texttt{MAG\_AUTO}$^{a}$    & \texttt{FLUX\_RADIUS}$^{b}$      & \multirow{2}{*}{$N_{S}^c$}      & \multirow{2}{*}{$N_{A}^d$} \\
                            &       (mag)           &       (arcsec)             &                               &                          \\     
\hline
F606W                       &    $[1.2,1.7]$        &       $[20,26]$            &           677                 &            498           \\
F814W                       &    $[1.2,1.7]$        &       $[19,25]$            &           654                 &            431           \\
\hline
\multicolumn{5}{l}{$^a$ Kron-like (Kron 1980) elliptical aperture magnitude.} \\
\multicolumn{5}{l}{$^b$ Estimated radius of the circle centered on the light barycenter} \\
\multicolumn{5}{l}{\ \ \ enclosing half the total flux.} \\
\multicolumn{5}{l}{$^c$ Number of sources selected for the PSF model ({\it red points in Fig. \ref{fig:Init_Select}})} \\
\multicolumn{5}{l}{$^d$ Number of sources accepted by PSFEx.}\\
\end{tabular}
\end{center}
\end{table}

\begin{figure*}
\begin{center}
    \includegraphics[width=0.49\textwidth]{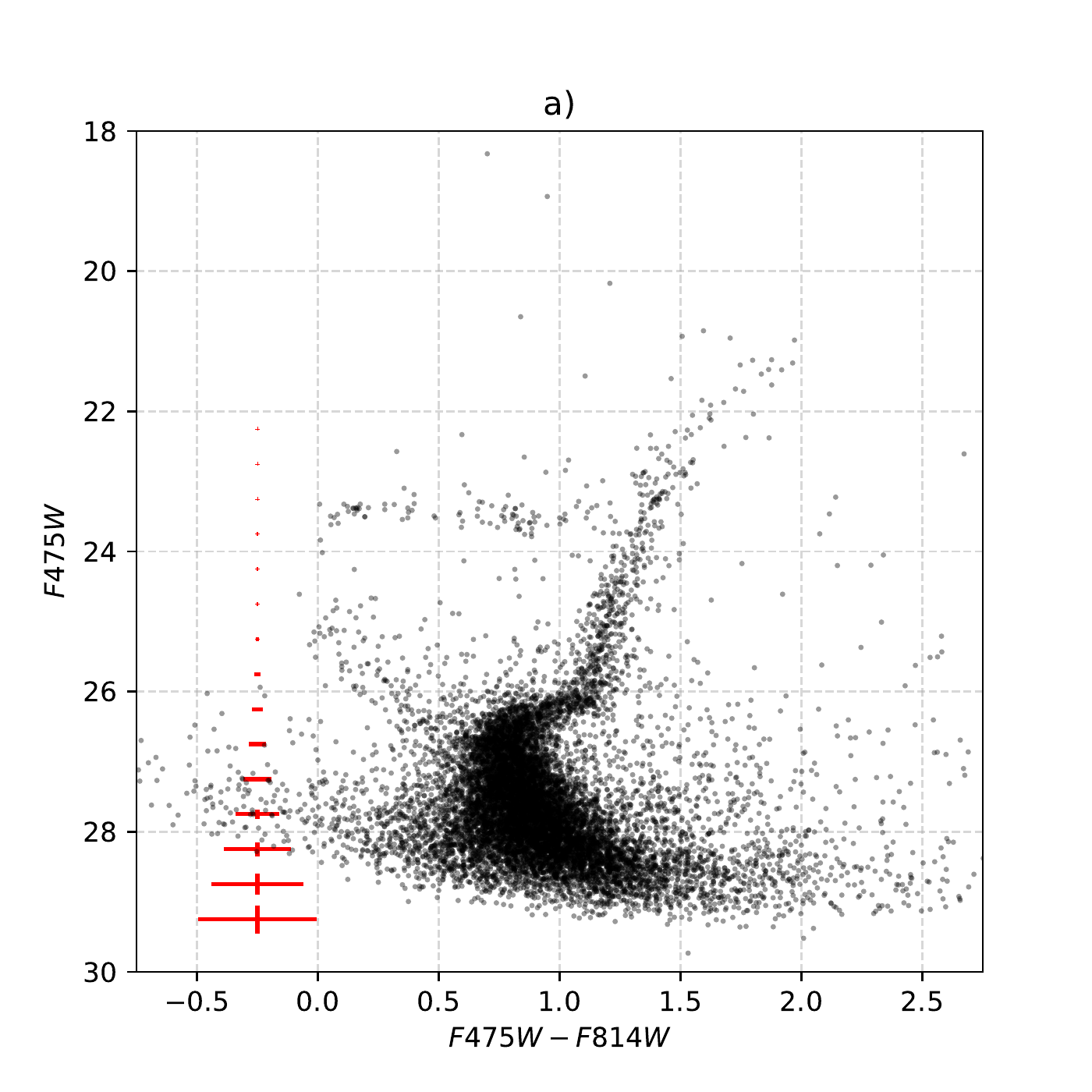}
    \includegraphics[width=0.49\textwidth]{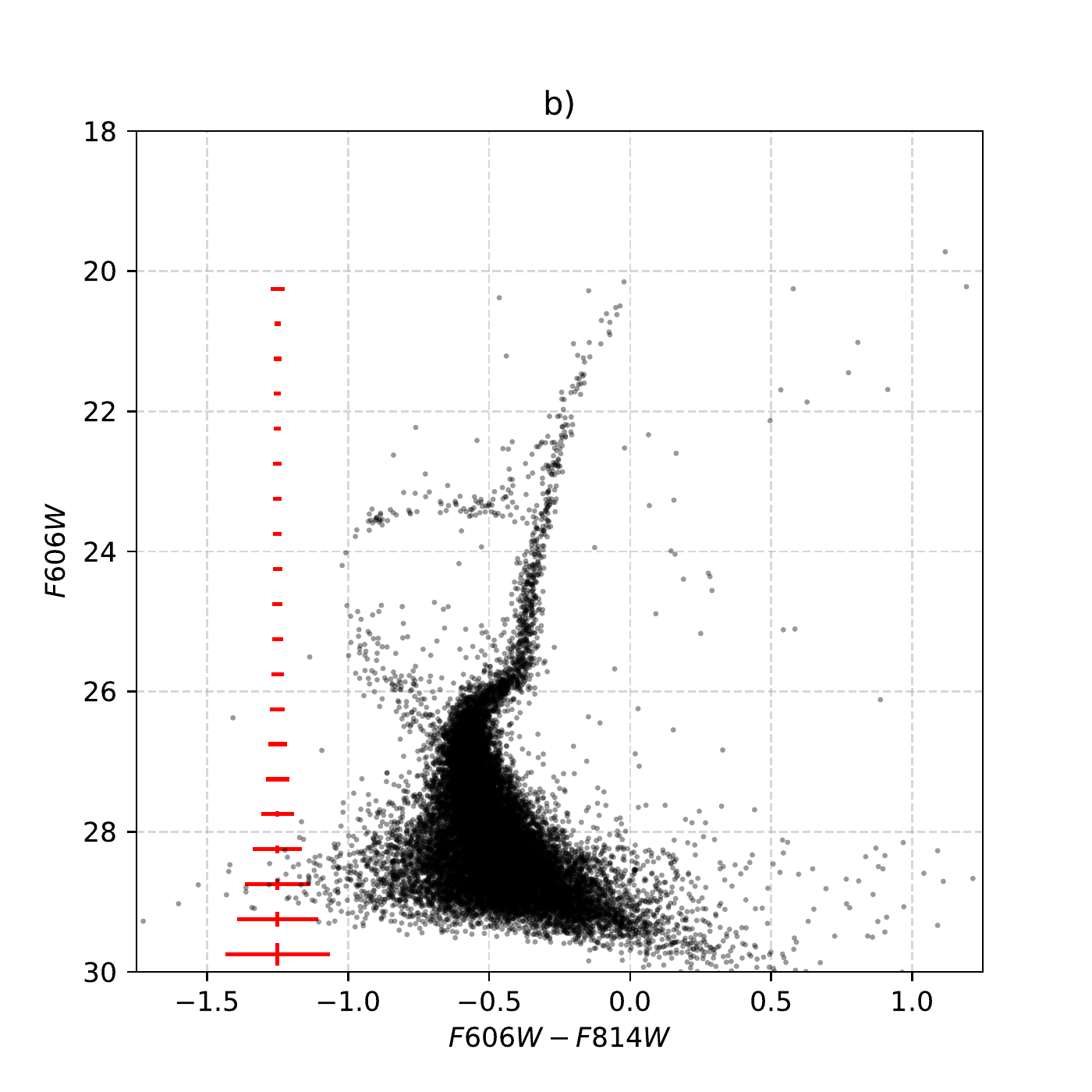}
    \caption{\label{fig:eri_CMDs}
    {\it (a)} F475W\,vs.\,F475W\,-\,F814W CMD (Vega mag) for 12,204 stars in the G16 data set.
    {\it (b)} F606W\,vs.\,F606W\,-\,F814W CMD (STmag) for 18,070 stars in the S21 data set.
    }
\end{center}
\end{figure*}

\begin{figure}
\begin{center}
        \includegraphics[width=0.7\columnwidth]{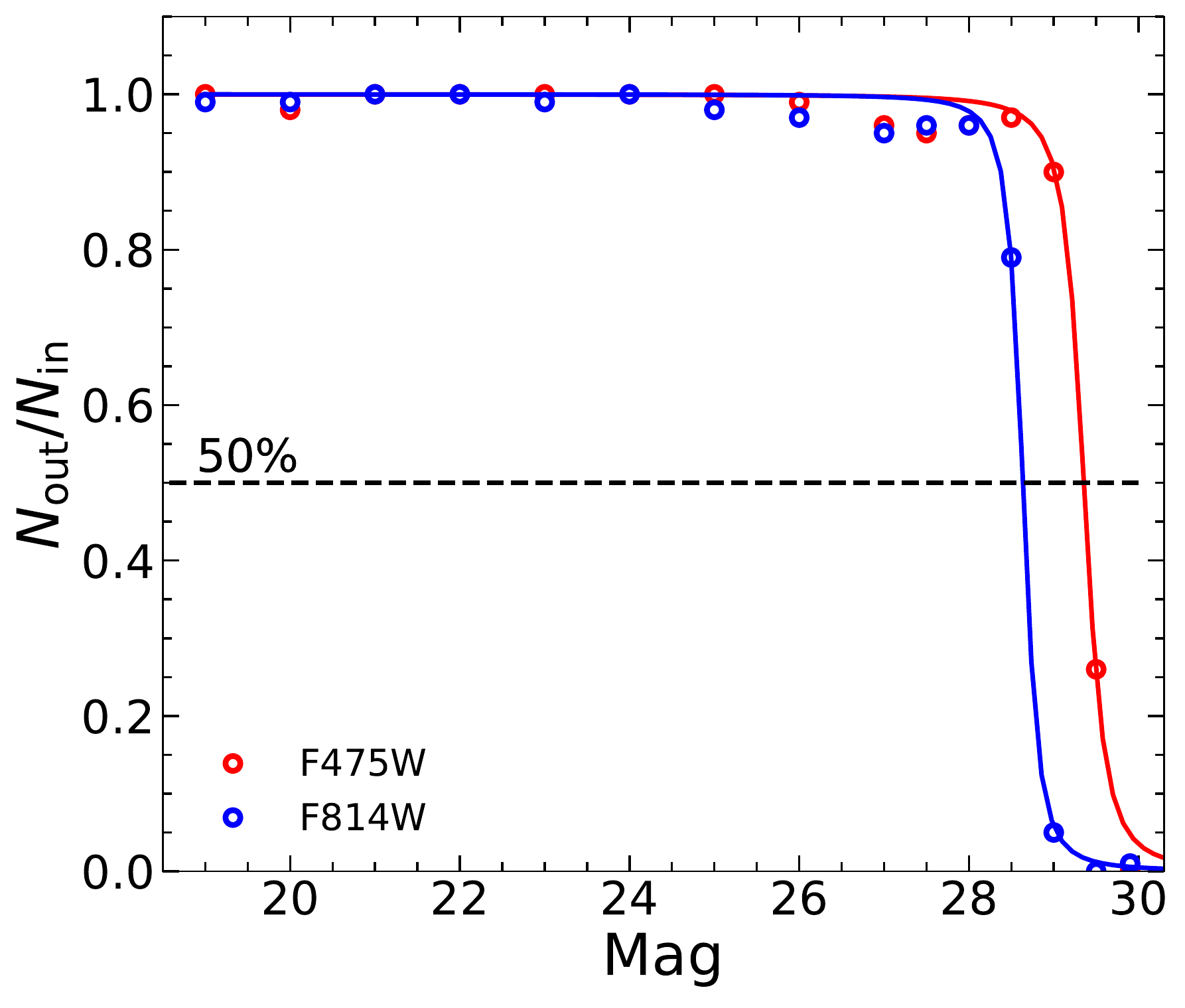}
        \caption{\label{fig:f475w_comp}
       Completeness of our Eri II photometry. The vertical axis shows the ratio $N_{\rm out}/N_{\rm in}$ between the number of recovered and inserted artificial stars. 
       The {\it red} and {\it blue open dots} show this ratio for stars detected according to our PSF model.
       The {\it solid lines} correspond to the fitted Pritchet function \citep{prit:90}.
       The photometry is 90\% complete at F475W\,=\,28.5 and F814W\,=\,29.0.
}
\end{center}
\end{figure}

Before the photometric analysis, the two images available for each filter were combined using the routine IRAF/IMCOMBINE.
PSF photometry was performed on each output image independently using the SExtractor\footnote{\url{http://www.astromatic.net/software/sextractor}} and PSFEx\footnote{\url{https://www.astromatic.net/software/psfex}} software packages \citep{bertin:96,bertin:11}. 
The latter is applied to build PSF models of the sources within a given image. SExtrator performance improves using PSF modeling.
The source detection over the output images was done according to \citet{rosa:17}, summarized in the following steps:
1) an initial photometric catalogue is generated using SExtrator;
2) a set of sources are manually selected as shown in Figs.\,\ref{fig:Init_Select}a and \ref{fig:Init_Select}b;
3) PSFEx uses this selection to build the PSF model; and
4) a final source catalogue is produced running SExtractor a second time.
Table \ref{tab:param_sel} defines the selection function. Objects which satisfy these conditions are with high probability non-spurious sources.

Cross-matching the sky position of the sources and selecting those with FLAGS\,$\leq$\,4 we obtain a sample of 12,159 stars with photometry in the two bands. SExtractor assigns FLAG\,=\,0 to well extracted sources and FLAGS\,=\,1,\,2,\,4 to sources affected by crowding issues\footnote{\url{https://sextractor.readthedocs.io/en/latest/index.html}}. 
We include sources flagged 1,\,2,\,4 to prioritize sample completeness.
The resulting sample is shown as a CMD in Fig.\,\ref{fig:eri_CMDs}a. 
The magnitudes from the G16 observations are in the Vega system.


\subsubsection{Sample completeness}

To determine the completeness of our photometry, we follow the {\it add and recover artificial star} method used by, e.g., \cite{rosa:17, mclau:94, prit:90}. We distribute artificially generated stars uniformly over the Eri II image, inserting 100 stars per magnitude interval in the range from 19 to 29 mag. We then repeat the extraction process described above.
The number of artificial stars recovered up to a given magnitude defines the completeness limit for that magnitude. 
Fig.\,\ref{fig:f475w_comp} shows our completeness function for the artificial stars using the Eri II PSF model.
The sample is $\approx$\,90\% complete at F475W\,$=$\,28.5 and F814W\,$=$\,29.0.
G21 do not mention explicitly their completeness limits but their CMD extends to F814W\,$=$\,28.5.
Table \ref{tab:samp_comp} summarizes the completeness of the samples used in this paper.

\begin{table}
\begin{center}
\caption{\label{tab:samp_comp}Sample completeness.}
\begin{tabular}{clcccc}
\hline
\multirow{2}{*}{Data set} &  \multirow{2}{*}{System}  &  \multicolumn{3}{c}{90\% completeness}    & \multirow{2}{*}{N$_*^a$}  \\
                    &                           &        F475W  &  F606W  & F814W             &                             \\
\hline
G16                 &          Vega             &        29.0   &   --    &  28.5             &          10,670              \\
S21                 &          STmag            &         --    &  28.7   &  29.2             &          13,493              \\
\hline
G21                 &          Vega             &         --    &   --    &  28.5             &            --               \\
\hline
\multicolumn{6}{l}{$^a$ Number of stars defining 90\% completeness.} \\
\end{tabular}
\end{center}
\end{table}

\section{Methodology}\label{sec:method}

The SFH of resolved stellar systems can be traced through isochrone fitting to their CMDs.
Initially, a common method to fit theoretical isochrones to CMDs was based on subjective visual criteria.
In the last twenty years objective quantitative methods based on $\chi^2$ minimization, maximum likelihood or Bayesian inference have become possible.
The latter allow to estimate not only the physical parameters describing the stellar population but also their uncertainties.
The case of galaxies with several bursts of star formation but poorly populated CMDs is particularly challenging since the number of free parameters is large enough
to make binning methods unappealing \citep{ramirez:19}.

In this paper we use the Bayesian hierarchical method developed by A21 to infer the SFH of resolved stellar populations. The SFH is expressed as a linear combination of $N_i$ stellar populations, each one described by an isochrone of known age and metallicity \citep{Dolphin:97,Small:13}. From the posterior probability distribution function (PDF) we obtain the relative contribution $a_i$ of the $i^{th}$ isochrone with its corresponding confidence interval. Since data binning is not required, this method is well suited to study populations with small numbers of stars, down to a few hundreds.

In a Bayesian hierarchical scheme we have two kind of parameters: population parameters, like $a_i$ (proportional to the star formation rate, SFR) and $\phi$ (the initial mass function, IMF), and individual 
parameters, like $M_{j}^{k}$, the absolute magnitude of the $j^{th}$ star in the $k$ band according to our isochrones. 
$k\,=\,1,\,2$ correspond to the F475W and F814W bands, respectively, in the case of the G16 data set, or to the
F606W and F814W bands, respectively, for the S21 sample.
The population parameters determine the PDF of the individual parameters. 
The Galactocentric distance $r_{Eri}$ and the extinction $A_k$ are taken as fixed quantities, as well as the set of isochrones which determine $M_{i}^{k}$. 
The locus described by each isochrone in the CMD is parametrized as a function of the stellar mass $m$. 
$M_i^k$ is then a unique function of $m$ for the $i^{th}$ isochrone in the $k^{th}$ band.
We model the absolute magnitude of a star of mass $m$ as a random variable of a convoluted isochrone such that $M_{j}^{k}\sim\mathcal{N}(\mathcal{M}_{i}^{k}(m), \sigma_{i}^{k})$. For simplicity, the dispersion $\sigma_{i}^{k}$ is taken as a constant, i.e., the isochrone is dispersed uniformly near each value of $m$.
The posterior PDF of the population parameters given the data $F_j^k$ with error $e_k^k$ follows then from Bayes theorem (A21)
\begin{equation}\label{eq:bayes}
    {\rm P}(\bm{a} \vert F_{j}^{k}) \propto {\rm P}(\bm{a}) \prod_{j=1}^{N_{D}}\int{ \frac{S(F_{j}^{k}){\rm P}(F_{j}^{k} \vert F_{{\rm true}}^{k}) \ {\rm P}(M_{j}^{k}\vert \bm{a})}{\ell(\bm{a},S)} } dM_{j}^{k},
\end{equation}
where $S(F^k_j)$ is the completeness function, a constant for a complete sample, and ${\rm P}(F_{j}^{k} \vert F_{{\rm true}}^{k})$ is the likelihood function, 
taken as the normal distribution
\begin{equation}
    \mathcal{N}(F_{j}^{k} \vert F_{{\rm true},j}^{k},e_{j}^{k}) \propto \exp{\left(\frac{F_{j}^{k}-F_{\rm true}}{e_{j}^{k}}\right)^2}.
\end{equation}
$F_{true}^{k} = M_{j}^{k}+\mu$ is the true apparent magnitude, an ideal measurement with no uncertainty, which is always unknown but it is needed for the statistical model structure \citep{luri:18}. $\mu$ is the distance modulus.

In Eq.\,(\ref{eq:bayes}) the denominator $\ell(\bm{a},S)$ is a normalization constant, and the hyper-prior ${\rm P}(\bm{a})$, 
where $\bm{a}=\{a_{i=1,...,N_{I}}\}$, $a_i\,>\,0$ and $\sum_{i}a_i=1$, is the symmetric Dirichlet distribution \citep{Walms:13}. 
The prior ${\rm P}(M_{j}^{k}\vert \bm{a})$ is defined in A21 as
\begin{equation}\label{prior_Mf}
P(M_{j}^{k}\vert \textbf{a})\,\propto\,\sum_{i=1}^{N_{iso}}\,a_{i}\,\int_{m_{l,i}}^{m_{u,i}}\phi(m)\,\prod_{k=1}^{3} \mathcal{N}(M^{k}_{j}\vert \mathcal{M}^{k}_{i}(m),\sigma_i^k)dm.
\end{equation}
In Eq.\,(\ref{prior_Mf}) the IMF $\phi(m)$, the lower and upper mass limits ($m_l,m_u$), and the isochrones enter as fixed quantities. See A21 for details.

For a complete sample the distribution of the number of stars along the MS is determined by the IMF.
If the sample is complete to a limiting magnitude, it is necessary to renormalize the statistical model to compensate for the missing stars.
If we truncate the sample to this limiting magnitude, the integral in Eq.\,(\ref{eq:bayes}) must be truncated to this limit.
In this way we ensure that the posterior PDF is correctly normalized according to the number of stars in the IMF.

\subsection{Stellar tracks and isochrones}

We use two independent sets of isochrones for our Bayesian inference.

\subsubsection{PARSEC isochrones}

The PARSEC isochrones\footnote{\url{http://stev.oapd.inaf.it/cgi-bin/cmd}} \citep{Bres:12} are available for different stellar mass ranges depending on the stellar metallicity $Z$.
For $0.0001$\,$\le$\,$Z$\,$\le$\,$0.02$ the isochrones extend from 0.1 to 350\,\msun;
for $0.03$\,$\le$\,$Z$\,$\le$\,$0.04$ from 0.1 to 150\,\msun;
and for $Z$\,=\,$0.06$ from 0.1 to 20\,\msun.
We use their $[\alpha/{\rm Fe}]=0$ (scaled-solar) isochrones, which include overshooting, atomic diffusion and the parameter $\eta$ describing the mass loss rate in the red giant branch (RGB) is set to $\eta=0.2$.

\subsubsection{BaSTI isochrones}

The BaSTI isochrones\footnote{\url{http://basti-iac.oa-abruzzo.inaf.it}} \citep{Hidalgo18} are available for $0.00001$\,$\le$\,$Z$\,$\le$\,$0.03905$ covering the mass range from 0.1 to 15\,\msun. 
We use their scaled-solar isochrones, including overshooting, atomic diffusion, and $\eta=0.3$.

\subsection{Isochrone grids}

From the BaSTI and PARSEC isochrone libraries, we assemble the two grids of isochrones listed in Table\,\ref{tab:iso} that will be used
below to infer the age-metallicity distribution of Eri II.
We remark here that the $Z$\,=\,$0.00001$ and $0.00005$ isochrones, {\it available} in the BaSTI data set, {\it are not available} for the PARSEC set.

\begin{table*}
\caption{\label{tab:iso}Isochrone Grids}
\begin{tabular}{ccccccccccc}
\hline
\multirow{2}{*}{Grid} & \multirow{2}{*}{Isochrones}    & Age       &   Step                   & \multicolumn{6}{c}{$Z$}                     & \multirow{2}{*}{N$_{iso}$} \\
                      &                                & (Gyr)     &   (Gyr)     & 0.00001 & 0.00005 & 0.0001 & 0.0002 & 0.0005 & 0.001     &                            \\     
\hline
A                     &  BaSTI                         & [1,14]    &   0.5       &  \chmk  &  \chmk  & \chmk  & \chmk  & \chmk  &           &            135             \\
\hline
B                     &  PARSEC                        & [1,14]    &   0.5       &         &         & \chmk  & \chmk  & \chmk  & \chmk     &            108             \\
\hline
\end{tabular}
\end{table*}

\begin{table}
\begin{center}
\caption{\label{tab:bic}Information Criteria.}
\begin{tabular}{cccccc} 
\hline
\multirow{2}{*}{Figure} & \multirow{2}{*}{Data} & \multirow{2}{*}{Grid}    & Maximum       &   \multirow{2}{*}{BIC}  &   \multirow{2}{*}{AIC} \\
                        &                     &                          & Likelihood    &                         &                        \\
\hline
{\it 5a}  & G16      &     A    &       -12669     &   26592   &     25608   \\
{\it 5b}  & S21      &     A    &       -6537      &   14361   &     13345   \\
\\
{\it 5c}  & G16      &     B    &       -14943     &   30890   &     30102   \\
{\it 5d}  & S21      &     B    &       -6970      &   14970   &     14157   \\
\hline

\end{tabular}
\end{center}
\end{table}

\begin{figure*}
\begin{center}
    \includegraphics[width=\textwidth,height=0.19\textheight]{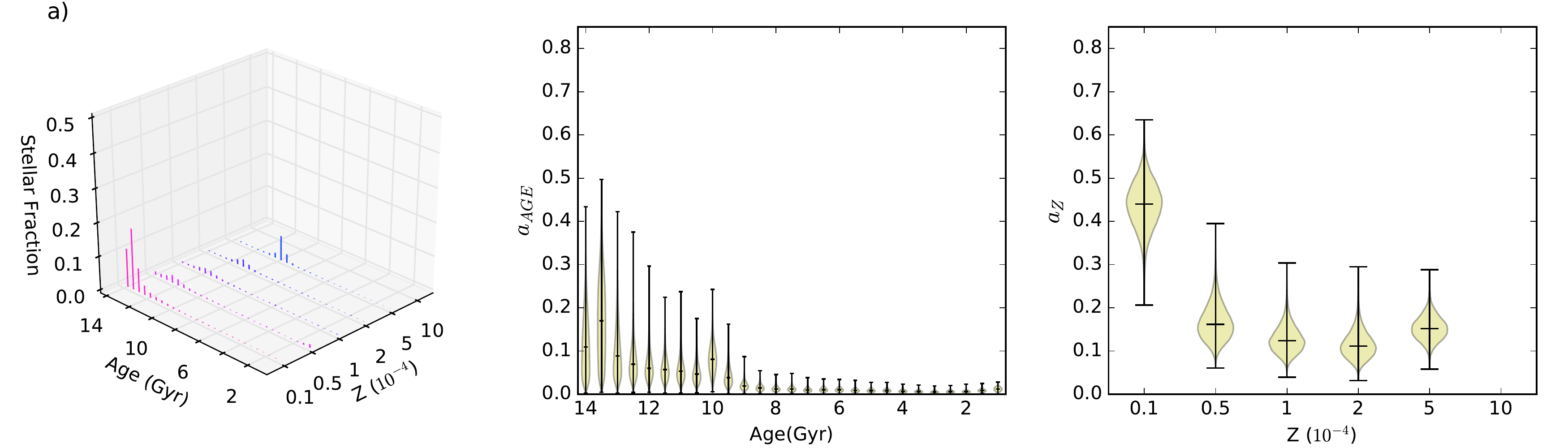}
    \includegraphics[width=\textwidth,height=0.19\textheight]{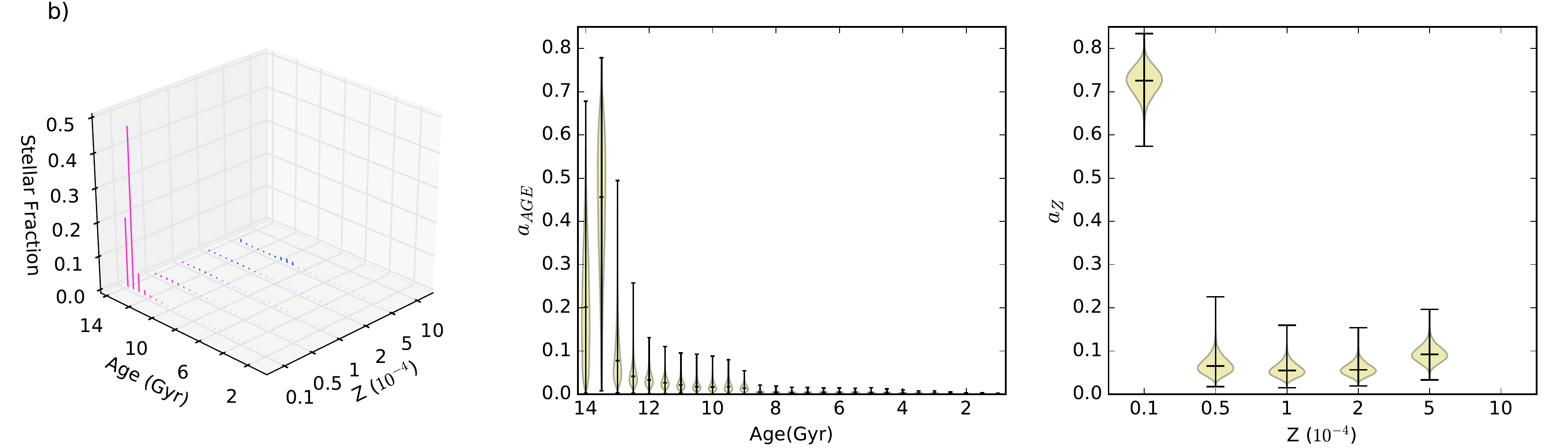}
    \includegraphics[width=\textwidth,height=0.19\textheight]{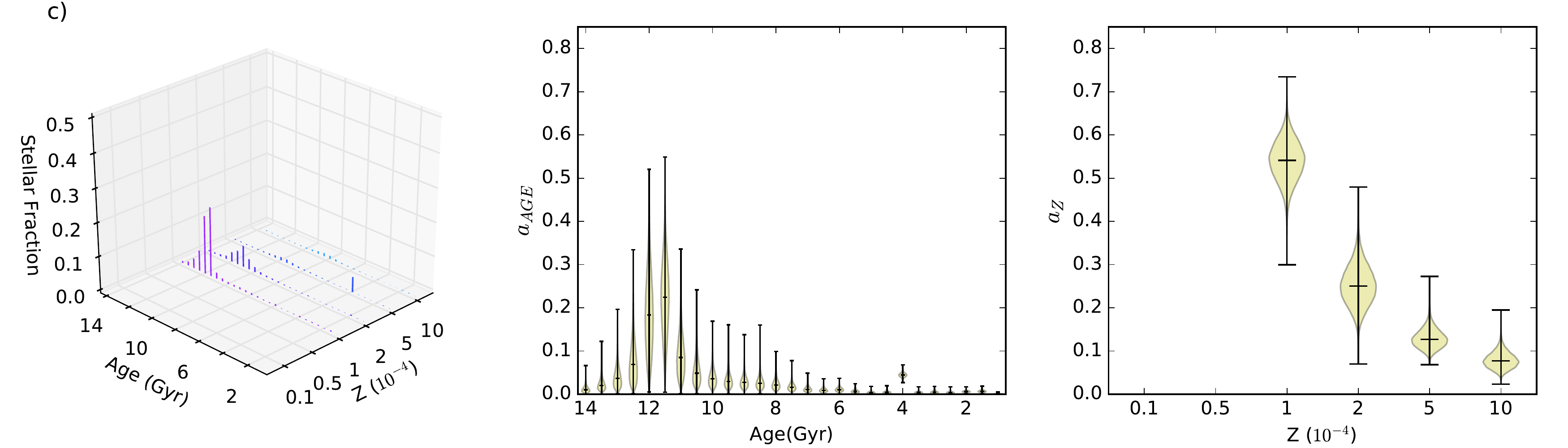}
    \includegraphics[width=\textwidth,height=0.19\textheight]{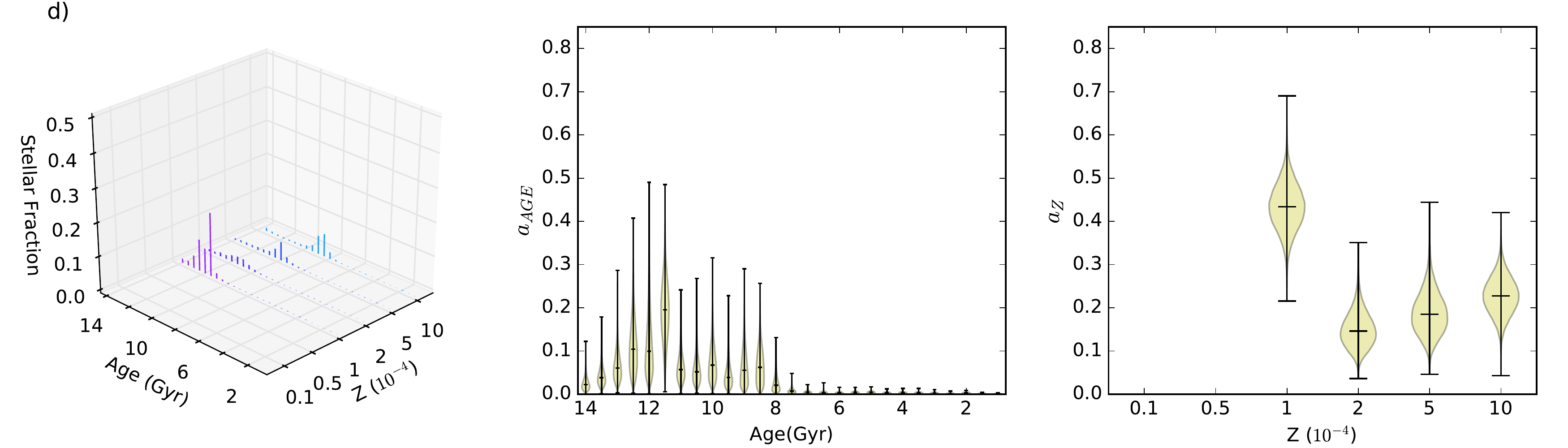}
    \caption{\label{fig:sfh_basti}
    AMD of Eri II inferred from the G16 and S21 data sets using isochrone grids A and B (Table\,\ref{tab:iso}).
    Our analysis is limited to the 90\% sample completeness regime.
    {\it (a)} G16 data set and isochrone grid A and $\sigma_i^k$\,=\,$0.05$\,mag.
    {\it (b)} S21 data set and isochrone grid A and $\sigma_i^k$\,=\,$0.03$\,mag.
    {\it (c)} G16 data set and isochrone grid B and $\sigma_i^k$\,=\,$0.05$\,mag.
    {\it (d)} S21 data set and isochrone grid B and $\sigma_i^k$\,=\,$0.03$\,mag.
    }
\end{center}
\end{figure*}

\begin{figure*}
\begin{center}
    \includegraphics[width=0.49\textwidth]{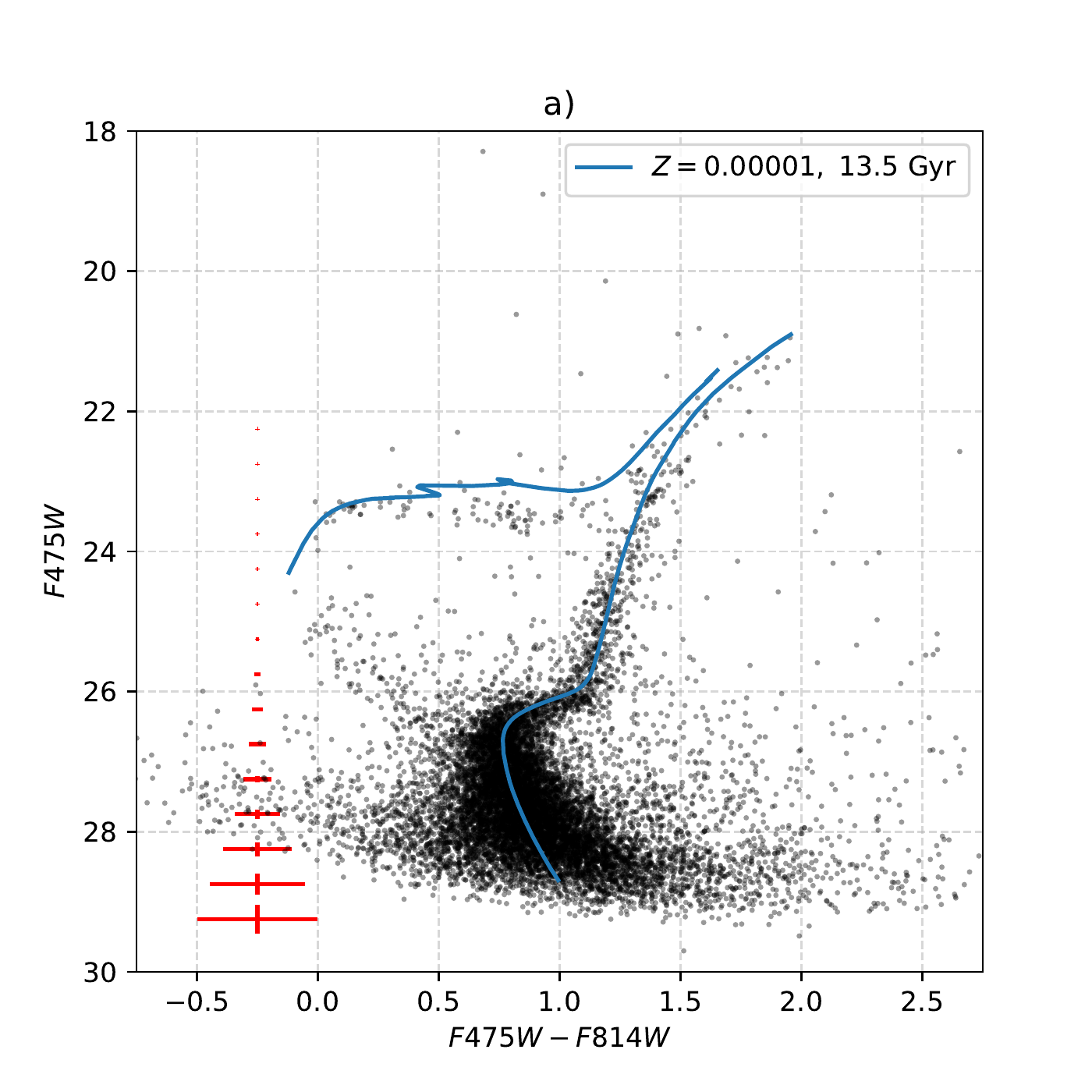}
    \includegraphics[width=0.49\textwidth]{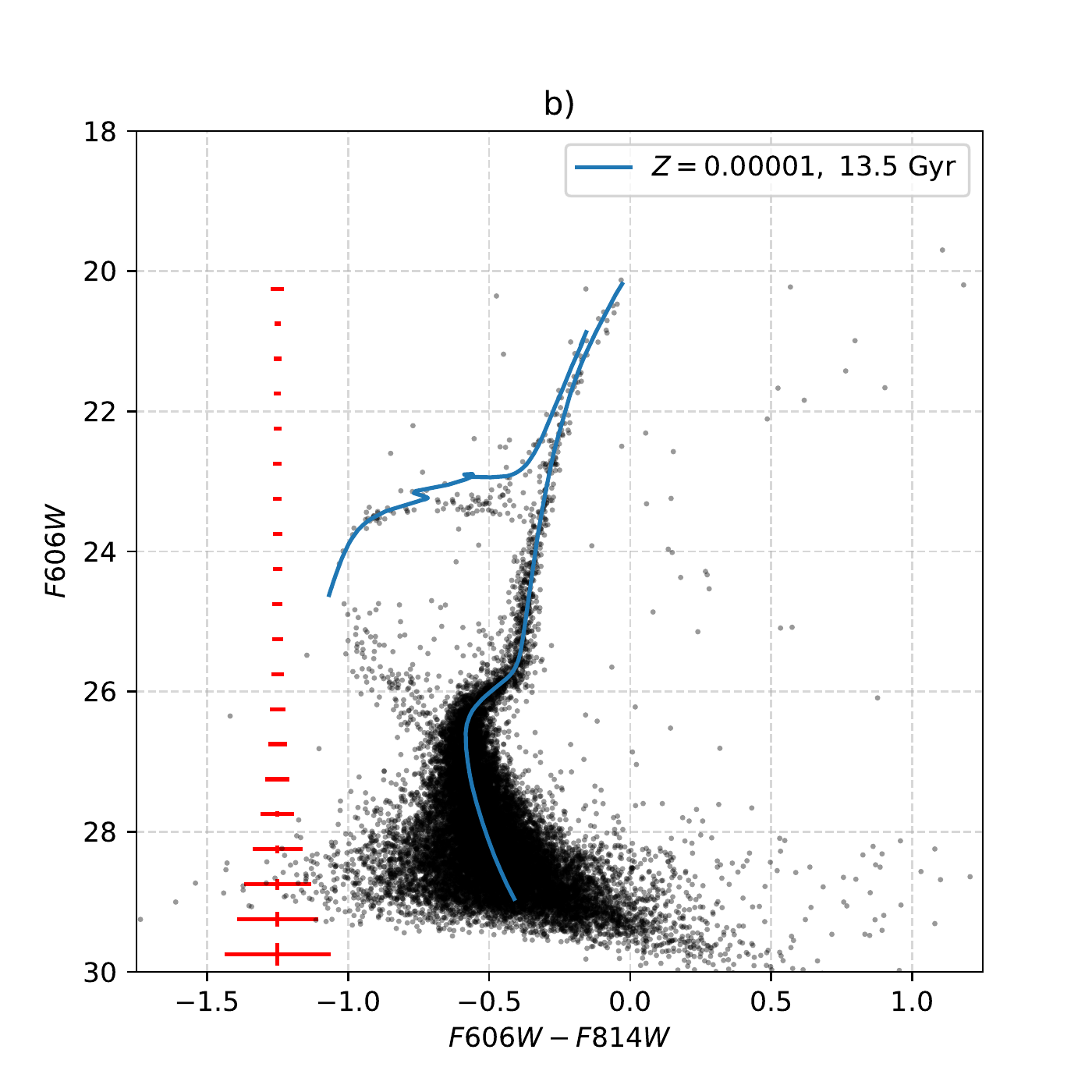}
    \includegraphics[width=0.49\textwidth]{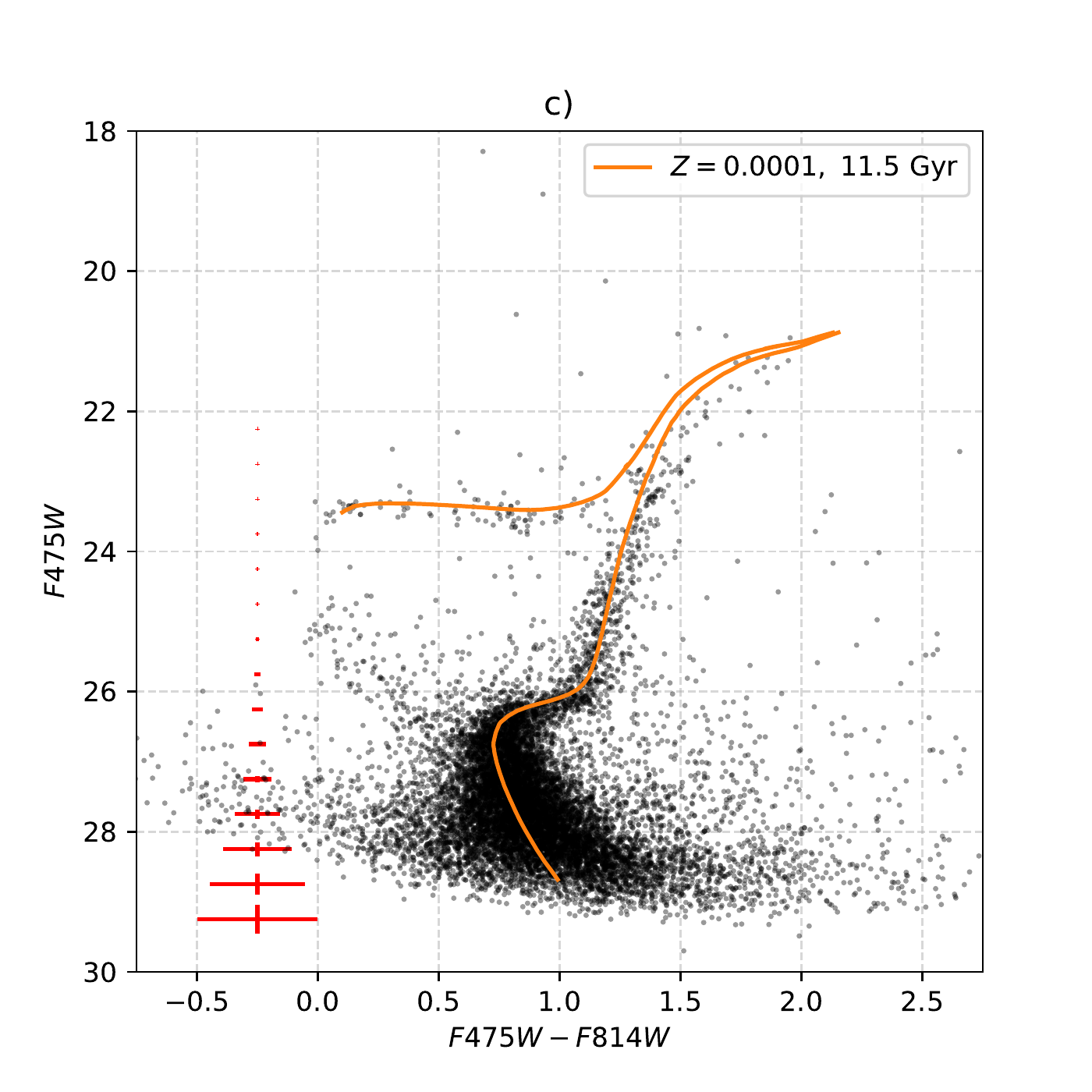}
    \includegraphics[width=0.49\textwidth]{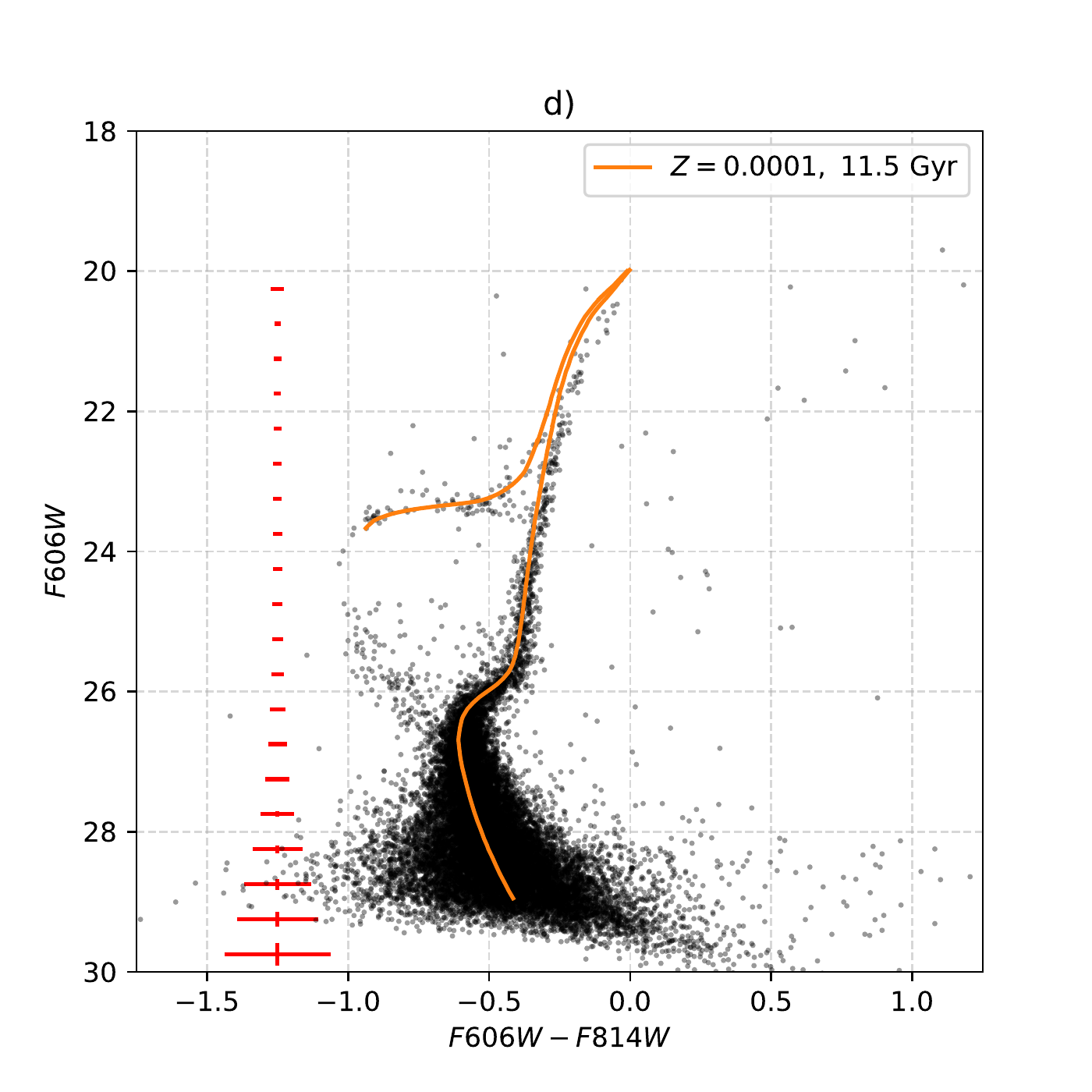}
    \caption{\label{fig:eri_ISOs}
    {\it (a,c)} F475W\,vs.\,F475W\,-\,F814W CMD (Vega mag) for 12,204 stars in the G16 data set.
    {\it (b,d)} F606W\,vs.\,F606W\,-\,F814W CMD (STmag) for 18,070 stars in the S21 data set.
    The 13.5\,Gyr, $Z$\,=\,$0.00001$ BaSTI isochrone, corresponding to the maximum value of $a_{AGE}$ in Figs.\,5a,b, is shown
    as a {\it blue} line in the {\it upper row} panels. The 11.5\,Gyr, $Z$\,=\,$0.0001$ PARSEC isochrone, corresponding to the maximum value of 
    $a_{AGE}$ in Figs.\,5c,d, is shown as an {\it orange} line in the {\it bottom row} panels.
    }
\end{center}
\end{figure*}
\begin{figure*}
\begin{center}
    \includegraphics[width=\columnwidth]{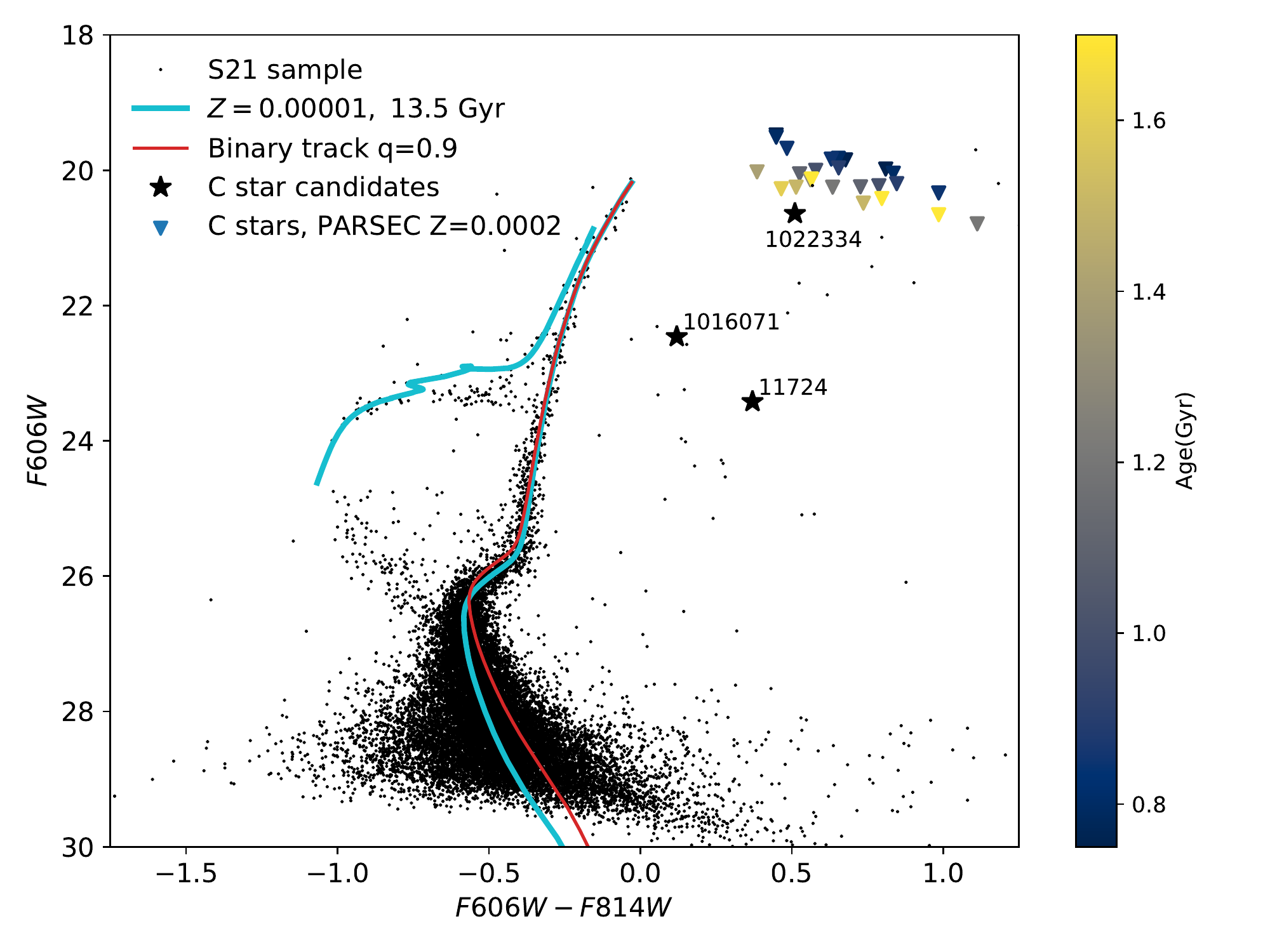}
    \includegraphics[width=\columnwidth]{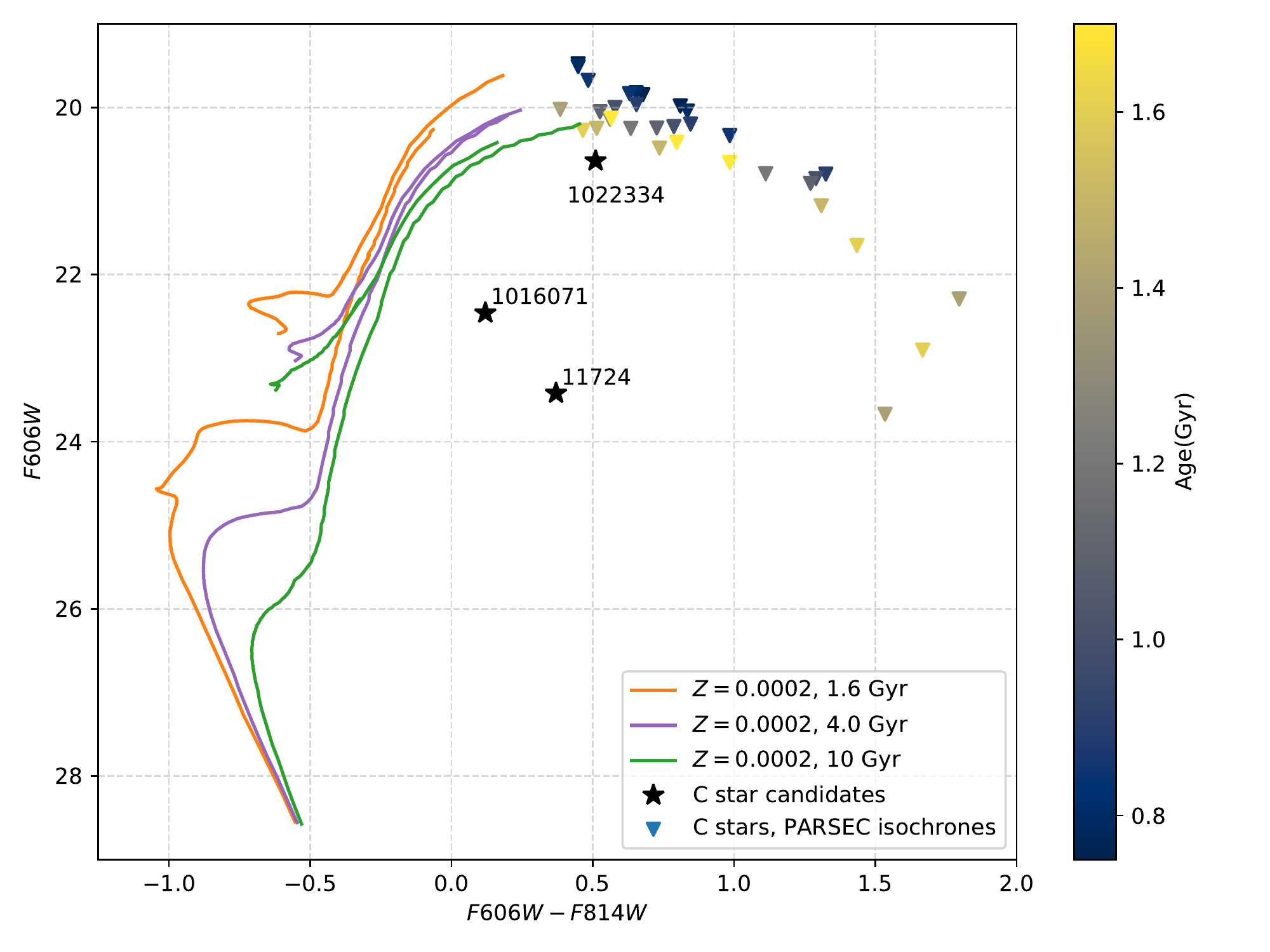}
    \caption{\label{fig:cstars}
     F606W\,-\,F814W vs. F606W CMD in STmag units indicating the position of the \citet{zout:20} C star candidates, and the C stars for
     the $Z\,=\,0.0002$ PARSEC isochrones. No photometric errors are provided by \citet{zout:20}. The {\it cyan} line in the
     {\it left-hand-side} panel
     is the 13.5\,Gyr, $Z$\,=\,$0.00001$ single star BaSTI isochrone, corresponding to the maximum value of $a_{AGE}$ in Figs.\,5b. 
     The red line in this panel corresponds to the expected position of this isochrone in the CMD when unresolved binary stars of mass 
     ratio $q$\,=\,$m_2/m_1$\,=\,0.9 are taken into account. Single star PARSEC isochrones are shown in the {\it right-hand-side} panel.
     }
\end{center}
\end{figure*}
\begin{figure}
\begin{center}
    \includegraphics[width=1.\columnwidth]{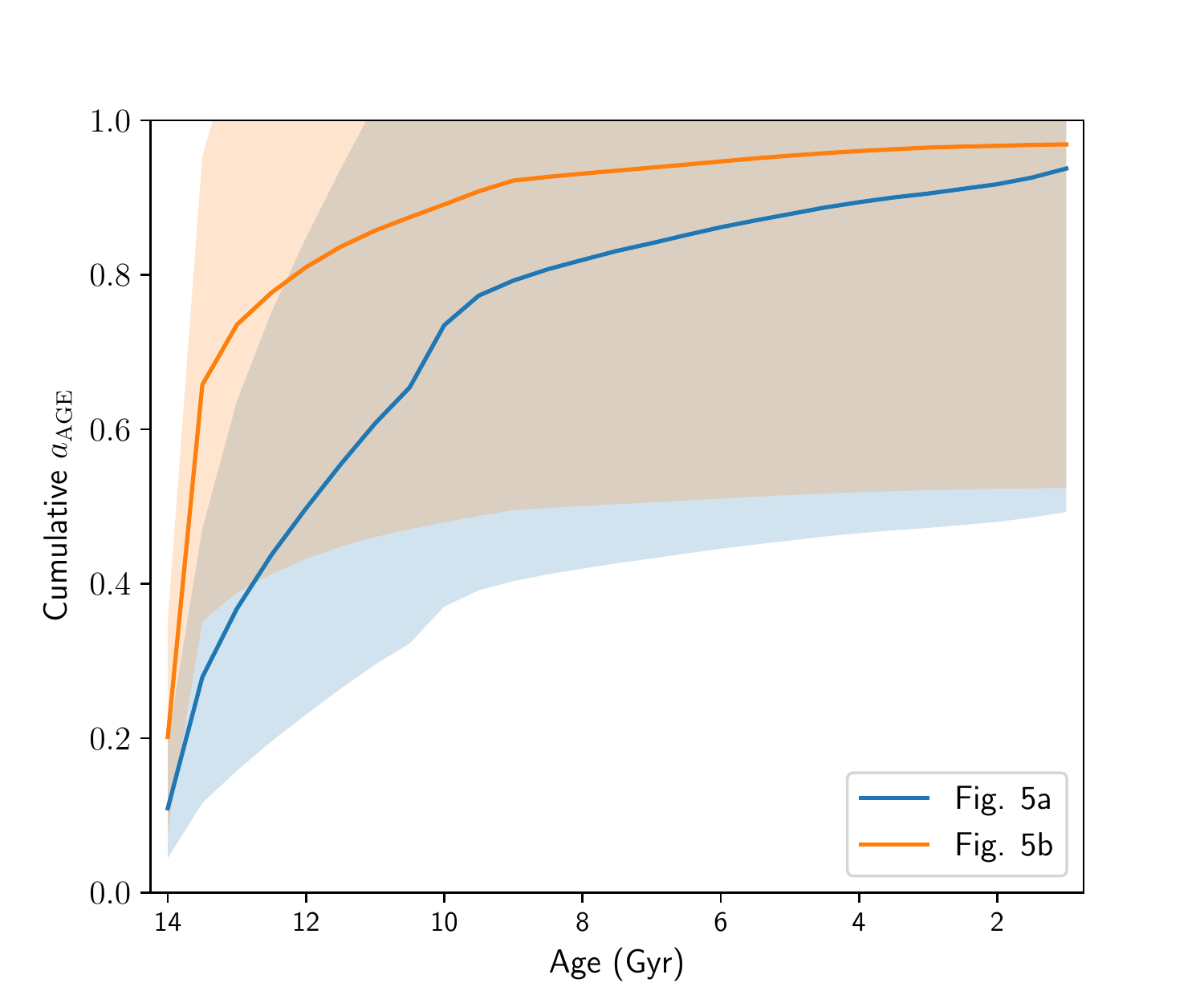}
    \caption{\label{fig:cumul}Normalized cumulative number of stars vs. age corresponding to the AMDs in Fig.\,5a ({\it blue} line, G16, isochrone
     Grid A) and Fig.\,5b ({\it orange} line, S21, isochrone grid A). The shaded areas indicate the percentiles 10 and 90 of the corresponding
     distribution.
     }
\end{center}
\end{figure}

\section{Age-metallicity distribution of Eri II}\label{sec:age_metal}

To infer the age-metallicity distribution (AMD) from the CMD we proceed as follows.
The isochrones listed in Table\,\ref{tab:iso} define the loci occupied in the CMD by stars of the corresponding age and metallicity .
The isochrones are parametrized by the stellar mass and then re-sampled such that the difference in magnitude between consecutive points along the isochrone is $\approx\,0.005$ mag.
To build the {\it posterior} PDF we sample the $a_i$ parameter space using a $10^5$ step Markov chain Monte Carlo (MCMC)
process. 
We use the {\it Stan} MCMC platform\footnote{\url{https://mc-stan.org}} due to its recognized reliability and acceptance of hard constraints, 
e.g., $a_i\geq0$ and $\sum a_i = 1$, as required in our case.
From this posterior we infer the distribution of values of $a_i$ and its confidence interval.
We assume the \cite{kr01} IMF.

In this paper we use $\mu$\,=\,$22.8$ for the distance modulus to Eri II, as reported by \cite{crnojevic:16}. 
From the NED\footnote{\url{https://ned.ipac.caltech.edu}} database we obtain the extinction in the direction of Eri II in the 3 bands in use 
$(A_{475},\,A_{606},\,A_{814})$\,=\,$(0.033,\,0.025,\,0.015)$ mag.

The AMDs inferred from the G16 and S21 Eri II samples (Section\,\ref{sec:data_reduction} and Fig.\,\ref{fig:eri_CMDs}) using isochrone
grids A and B (Table\,\ref{tab:iso}) are displayed in Fig.\,\ref{fig:sfh_basti}.
The height of the bars in the 3D plots {\it (left panels)} of Fig.\,\ref{fig:sfh_basti} is the median of the distribution of $a_i$ for the corresponding isochrone. 
The violin plots summarize the marginalized posterior PDF for age ({\it center panels}) and $Z$ {\it (right panels)}. 
The horizontal lines in each violin represent from {\it bottom} to {\it top} the 0, 50, and 100 percentiles of the distribution, respectively. 
The age-metallicity degeneracy \citep{worthey:94} is clearly noticeable in these distributions: the higher $Z$ PARSEC isochrones (Grid B) predict younger ages for the Eri II stellar population than the lower $Z$ BaSTI isochrones (Grid A).
We remind the reader that the $Z$\,=\,$0.00001$ and $0.00005$ isochrones {\it are not available} for the PARSEC set in Grid B.

The agreement between the AMD's in Figs.\,\ref{fig:sfh_basti}{\it a,b}
and Figs.\,\ref{fig:sfh_basti}{\it c,d} is quite remarkable given the different photometries
and isochrones in use. In Table\,\ref{tab:bic} we list the values for the Bayesian information criterion (BIC) and the Akaike information criterion (AIC) for the four inferences in Fig.\,\ref{fig:sfh_basti}.
These criteria\footnote{
$BIC\,=\,ln(n)\,k\,-\,2\,ln(\mathcal{L})$, where $\mathcal{L}$ is the maximum likelihood value, $n$ the number of data points, and $k$ the number of estimated free parameters.
$AIC\,=\,2k-2ln(\mathcal{L})$}
are partially based on the likelihood function and serve for model selection among a finite set of models \citep{gelman13}. Though BIC is always higher than AIC, the lower the value of these two measures, the better the model. Thus, the model
in \ref{fig:sfh_basti}{\it b} describes the data better than the model in \ref{fig:sfh_basti}{\it a}. Similarly, the model in \ref{fig:sfh_basti}{\it d} describes the data better than the model in \ref{fig:sfh_basti}{\it c}.
The similarities between the inferred AMD's for these two pairs of data and models support that the stellar evolution models play an important role. The solutions consistently compensate the lack of the most metal-poor models with younger ages.

In Fig.\,\ref{fig:eri_ISOs} we show the isochrones corresponding to the maximum $a_{AGE}$ in the AMDs of Fig.\,\ref{fig:sfh_basti}.
Comparing Fig.\,\ref{fig:eri_ISOs}b,d, it is apparent that the RGB is reproduced better by the BaSTI than by the PARSEC isochrones,
even though the opposite may be true for the horizontal branch (HB) stars. To a lesser degree, the same behaviour is observed when
comparing Fig.\,\ref{fig:eri_ISOs}a,c. For this reason {\it we consider the AMD in Fig.\,\ref{fig:sfh_basti}b to be the most realistic approximation to the true distribution}. From Fig.\,\ref{fig:cumul} we see that according to this AMD, $\approx$\,80\% of the stars in
Eri II are older than 13 Gyr. In contrast, the solution for the G16 data set (Fig.\,\ref{fig:sfh_basti}a) implies that $\approx$\,60\% of the stars in Eri II are younger than 13 Gyr. This discrepancy is likely due to the lower quality of the G16 (shorter exposures, lower signal-to-noise ratio, less deep images) compared to the S21 data set, and possibly indicates that we underestimated the photometric errors
for the G16 sample (see Section\,\ref{sec:data_reduction}).

\subsection{Carbon stars in Eri II}\label{sec:CS}

\cite{zout:20} identified three possible C stars in the field of Eri II.
One of them (ID1022334) was confirmed to be a C star and to belong to the bulk population of Eri II, the other two (ID11724,\,ID1016071) remain as candidates. 
The presence of C stars supports the existence of an intermediate age population in Eri II. 
According to the $Z\,=\,0.0002$ PARSEC evolutionary tracks\footnote{We do not have available this information for the BaSTI tracks.}, stars of MS mass in the range $2.2$ to $1.55$ M$_\odot$ become C stars at age from $0.75$ to $1.7$ Gyr old, respectively.
The age of C stars increases slowly with metallicity, reaching $\approx 2$ Gyr near $Z_\odot$.
Since we do not detect any residual star formation extending to $\approx 2$ Gyr, it is likely that the C stars evolve from lower mass progenitors that increased their mass through stellar fusions ({\it blue-stragglers}, seen in significant numbers in the CMDs).
This mechanism requires the star density in Eri II to be high enough for stellar collisions to be important.

In Fig.\,\ref{fig:cstars} we show the position of the \cite{zout:20} C star candidates in the CMD together with the C stars expected from the $Z$\,=\,$0.0002$
PARSEC isochrones for $0.75$\,$\leq$\,age\,$\leq$\,$1.7$ Gyr.
Only the confirmed candidate ID1022334 has the luminosity and colour corresponding to C stars.
The C star nature of the other two candidates seems doubtful. 
Part of the difference between models and observations in Fig.\,\ref{fig:cstars} may result from inaccuracies in the \cite{aringer2009} C star spectral models used to compute the expected C star colors. 
In this case we used the $Z$\,=\,0.1\,$\times$\,$Z_\odot$ models, which is the minimum metallicity for which models are available.
Taken at face value, the observed (F606W,\,F814W) = (20.64,\,20.13) STmag
place star ID1022334 at 500 to 400 kpc from us, $\mu$\,=\,$(23.5,\,23.05)$ mag, respectively, while the reported $\mu$ to Eri II is 22.83 mag \citep{crnojevic:16}.

\subsection{Dating the star cluster in Eri II}\label{sec:DGC}

We made several attempts to date the star cluster in Eri II as a separate entity, with no success. We selected all the stars inside the {\it dashed white circle} in Fig.\,\ref{fig:eri_img}. The number of cluster stars is too low ($\approx$\,300) for our Bayesian hierarchical model to work properly. Our grids contain $\sim$\,100 isochrones, so in this case we have $\approx$\,3\,stars per isochrone, which prevents a proper statistical treatment of the cluster population. In Appendix\,\ref{sec:appendix} we use simulated stellar populations to show that this is in fact the case.

\section{Conclusions} \label{sec:summary}

We have performed a detailed stellar population analysis of Eridanus II using our Bayesian hierarchical model \citep{alzate21} to infer with an acceptable level of statistical significance relevant information about its SFH.

We find convincing evidence that the bulk of the stars in Eri II are very old, with an age of  $13.5_{-1}^{+0.5}$ Gyr and quite metal poor, with $Z=0.00001$ (see the AMD in Fig.\,\ref{fig:sfh_basti}b). In agreement with S21, we found that the 70\% of the stars were formed 700 Myr after Big Bang (Fig.\,\ref{fig:cumul}). This result is consistent with the width at half maximum ($\sim$500 Myr) of the derived star formation rate profile of G21.

We did not succeed in determining the age of the star cluster as an independent entity due in part to the small number of visible stars
directly associated with the cluster. Nor we find any evidence of the presence of an intermediate age population. Cluster stars are expected
to be included in the stellar population that we analyzed with no indication of bimodal distributions in age or in metallicity.

The survival of the star cluster for over 13 Gyr inside the DM halo of Eri II favours the model of a flat (core) rather than a cuspy (NFW) central density profile \citep[][2020 in preparation]{lora:12,lora:13}.

The lack of recent star formation implies that mass pumping of lower mass MS stars through {\it blue-straggler} fusions is responsible of forming the massive progenitors of the C stars seen today in Eri II.

From simulated stellar populations we conclude that 
{\it (a)} we can recover correctly the age of stellar populations of the same metallicity born in different star formation episodes separated in time by more than the time resolution of the isochrone grid in use;
{\it (b)} nearly coeval populations of different metallicity can be characterized as long as the number of stars in the less massive population is significant; and
{\it (c)} the size of the photometric errors included in the statistical model must resemble the true errors.

\section*{Acknowledgements}

We thank the anonymous referee for the careful reading of our manuscript and for useful suggestions that improved the quality and scope of this paper.
We thank very specially Denija Crnojevich who kindly guided us through the Eridanus II data.  
GB and JAA acknowledge financial support from the National Autonomous University of M\'exico (UNAM) through grant DGAPA/PAPIIT IG100319
and from CONACyT through grant CB2015-252364. 
VL gratefully acknowledges support from the \mbox{CONACyT} Research Fellowship program. 
BCS acknowledges financial support through PAPIIT project IA103520 from DGAPA-UNAM.
The research in this paper is part of the PhD thesis of J. A. Alzate in the Universidad Nacional Aut\'onoma de M\'exico (UNAM) graduate
program in astrophysics. He thanks the support from the Instituto de Radioastronom\'ia and Astrof\'isica, its staff, and the Consejo
Nacional de Ciencia y Tecnolog\'ia (CONACyT) for the scholarship granted. 

This work is based on observations made with the NASA/ESA Hubble Space Telescope, and obtained from the Hubble Legacy Archive,
which is a collaboration between the Space Telescope Science Institute (STScI/NASA), the Space Telescope European Coordinating
Facility (ST-ECF/ESAC/ESA) and the Canadian Astronomy Data Centre (CADC/NRC/CSA).

\section*{Data availability}
The \cite{gallart16} Eridanus II data underlying this article are available in the 
{\it Hubble Legacy Archive}, Proposal ID 14224, 2016, (\url{https://hla.stsci.edu/hlaview.html}).
The \citet{simon:21} Eridanus II photometry used in this article is available in the {\it astro-ph}, arXiv:2012.00043, 2020 (\url{https://arxiv.org/abs/2012.00043}).
We make use of the PARSEC isochrones \citep[][\url{http://stev.oapd.inaf.it/cgi-bin/cmd}]{Bres:12},
the BaSTI isochrones \citep[][\url{http://basti-iac.oa-abruzzo.inaf.it}]{Hidalgo18},
the {\it Stan} MCMC platform (\url{https://mc-stan.org}),
the NED database (\url{https://ned.ipac.caltech.edu}),
and the \cite{bertin:96}
SExtractor (\url{http://www.astromatic.net/software/sextractor})
and \cite{bertin:11}
PSFEx (\url{https://www.astromatic.net/software/psfex}) software packages. 
\bibliographystyle{mnras}
\bibliography{references}


\appendix

\section{AMD recovery by Bayesian inference}\label{sec:appendix}

\begin{table*}
\caption{\label{tab:a1}Simulation parameters. Columns 2,\,3,\,4 correspond to the metallicity, age and number of stars brighter than F475W=28.5 for the older population,
respectively, while columns 5,\,6,\,7 are the younger population counterparts.}
\begin{tabular}{ccccccccccc}
\hline                                                                            
\multirow{2}{*}{Simulation}  &   &   & \multicolumn{3}{c}{Old population} &      &     &   \multicolumn{3}{c}{Young population}    \\
                             &   &   &   $Z$ & AGE (Gyr) & $N_{\rm old}$  &      &     &     $Z$ & AGE (Gyr) & $N_{\rm young}$     \\
\hline                                                                            
   1                         &   &   &    0.0001    & $12.25,12,11.75$    & 4253 &     &     &        0.0001  &  11.0  &  1721      \\
\hline                                                                            
   2                         &   &   &    0.0001    & $12.25,12,11.75$    & 4253 &     &     &        0.0001  &  11.5  &  1730      \\
\hline                                                                            
   3                         &   &   &    0.0001    & $12.25,12,11.75$    & 4253 &     &     &        0.0002  &  11.5  &  1700      \\
\hline                                                                            
   4                         &   &   &    0.0001    & $12.25,12,11.75$    & 4253 &     &     &        0.0001  &  11.5  &  292       \\
\hline                                                                            
   5                         &   &   &    0.0001    & $12.25,12,11.75$    & 4253 &     &     &        0.0002  &  11.5  &  252       \\
\hline
\end{tabular}
\end{table*}

\begin{figure*}
\begin{center}
    \includegraphics[width=0.45\textwidth]{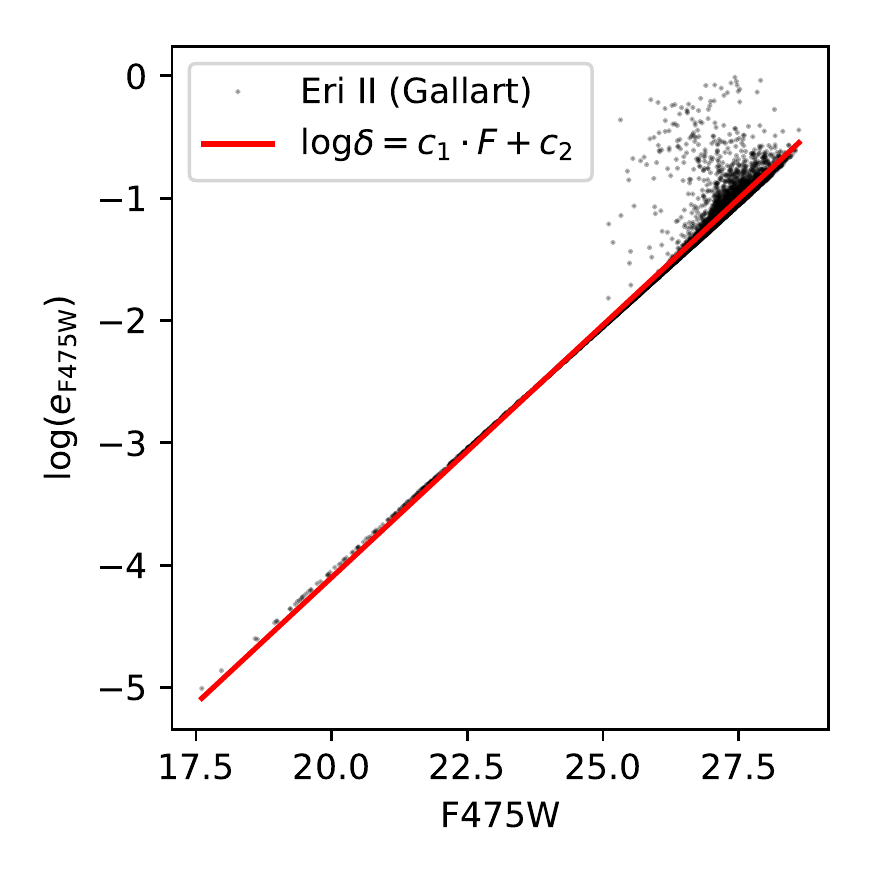}
    \includegraphics[width=0.45\textwidth]{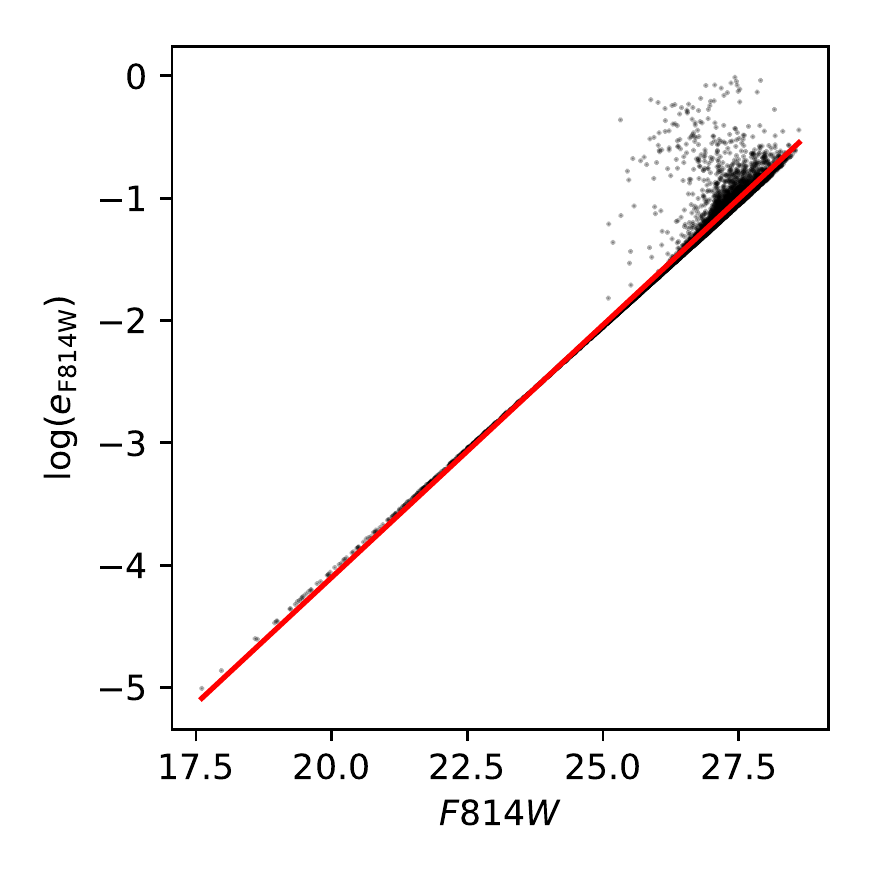}
    \caption{\label{fig:a11} Photometric errors for the G16 data sets. ({\it Left panel}) F475W, ({\it right panel}) F814W.
    The coefficients of the best fit ({\it red line}) are $(c1,\,c2)$\,=\,$(0.4171,\,-12.8011)$ and $(0.4129,\,-12.3588)$ for $\delta_{F475W}$ and $\delta_{F814W}$, respectively.
     }
\end{center}
\end{figure*}

\begin{figure*}
\begin{center}
    \includegraphics[width=0.45\textwidth]{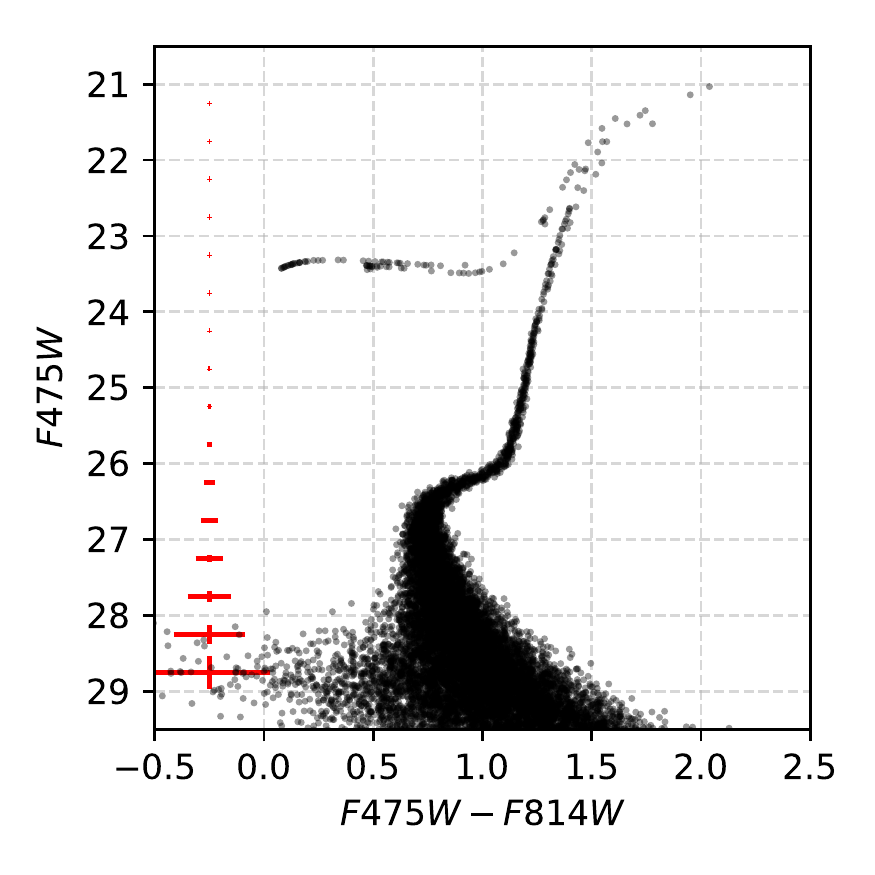}
    \includegraphics[width=0.45\textwidth]{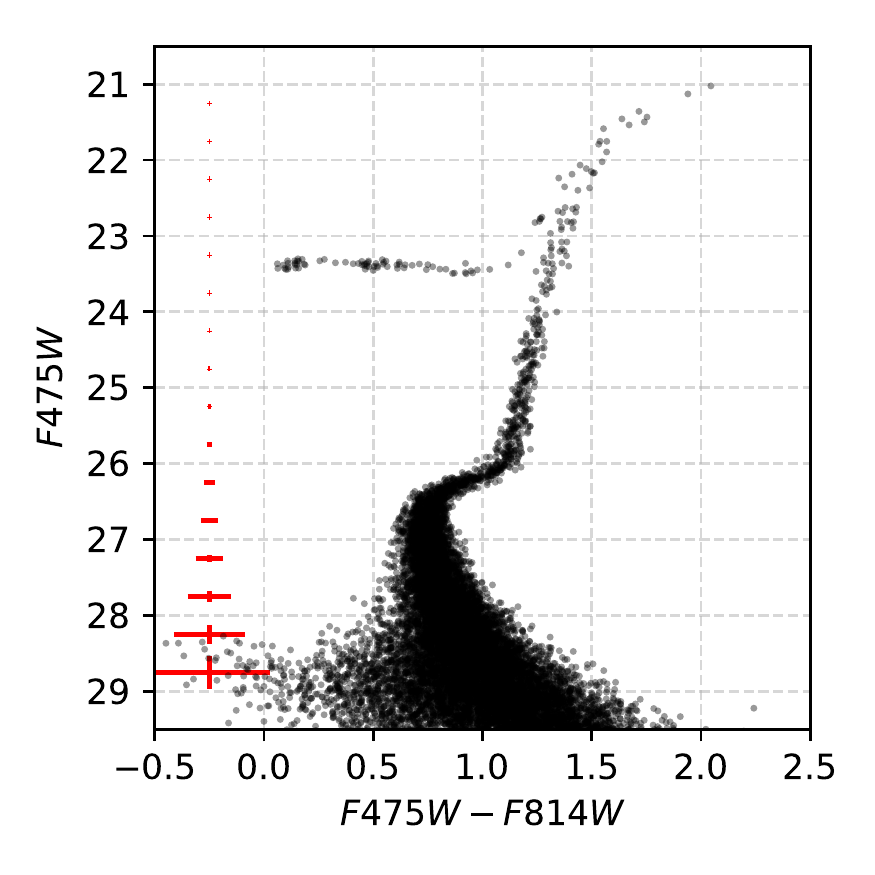}
    \caption{ \label{fig:a12} CMD for Simulation 3. ({\it Left panel}) Photometric errors from Fig.\,\ref{fig:a11}.
     ({\it Right panel}) Photometric errors increased by a constant value of 0.02 mag.
     }
\end{center}
\end{figure*}

\begin{figure*}
\begin{center}
    \includegraphics[width=\textwidth,height=0.16\textheight]{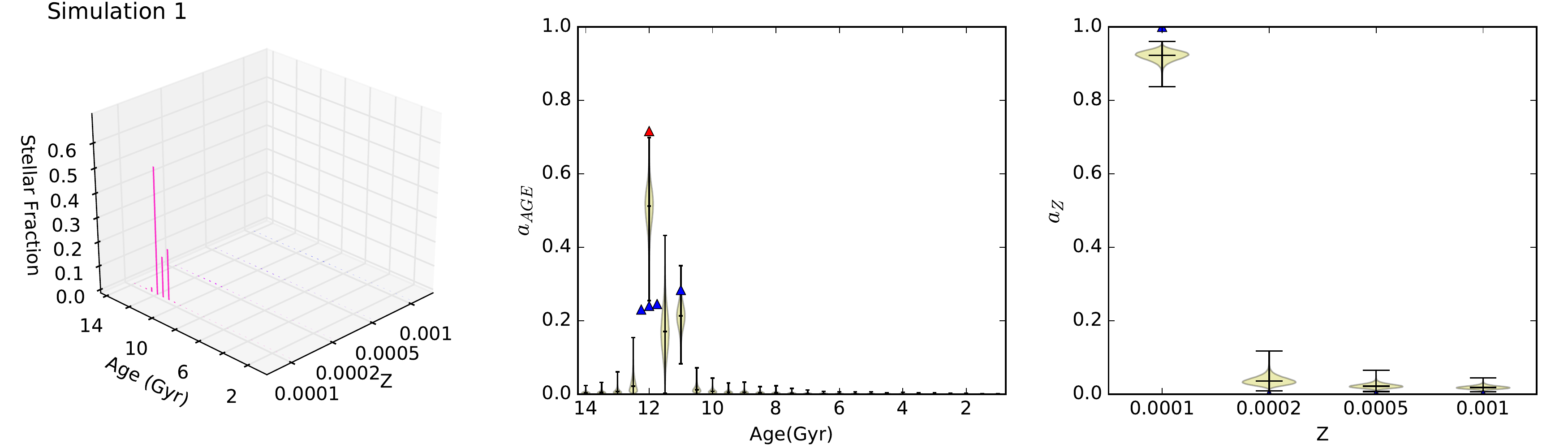}
    \includegraphics[width=\textwidth,height=0.16\textheight]{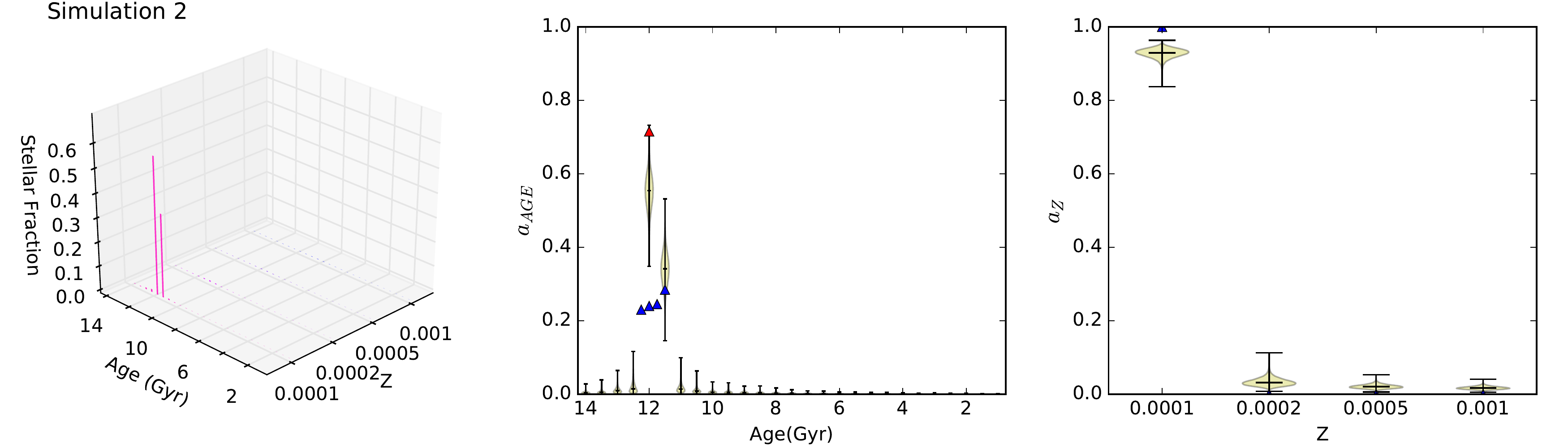}
    \includegraphics[width=\textwidth,height=0.16\textheight]{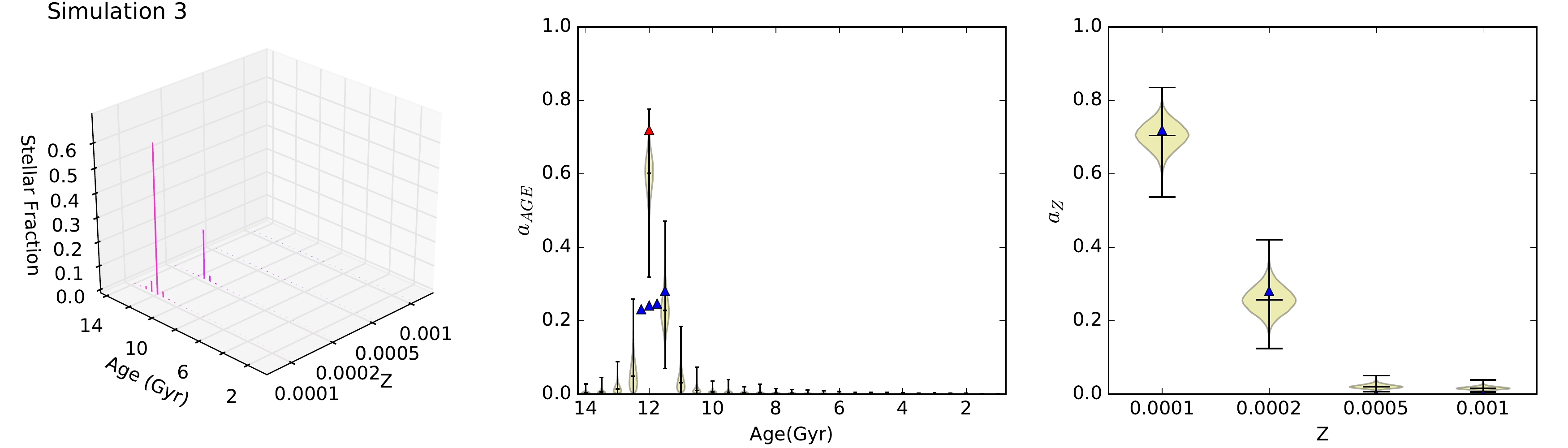}
    \caption{\label{fig:sfh_simu}
    Inferred AMD for Simulations 1 ({\it upper row}), 2 ({\it middle row}) and 3 ({\it bottom row}), derived using isochrone Grid B (Table\,\ref{tab:iso}) and $\sigma^{k}_{i}\,=\,0.01$ mag.
    The {\it blue triangles} represent the input values of $a_{AGE}$ and the {\it red triangle} the sum of $a_{AGE}$ for the older population.
    }
\end{center}
\end{figure*}

\begin{figure*}
\begin{center}
    \includegraphics[width=\textwidth,height=0.16\textheight]{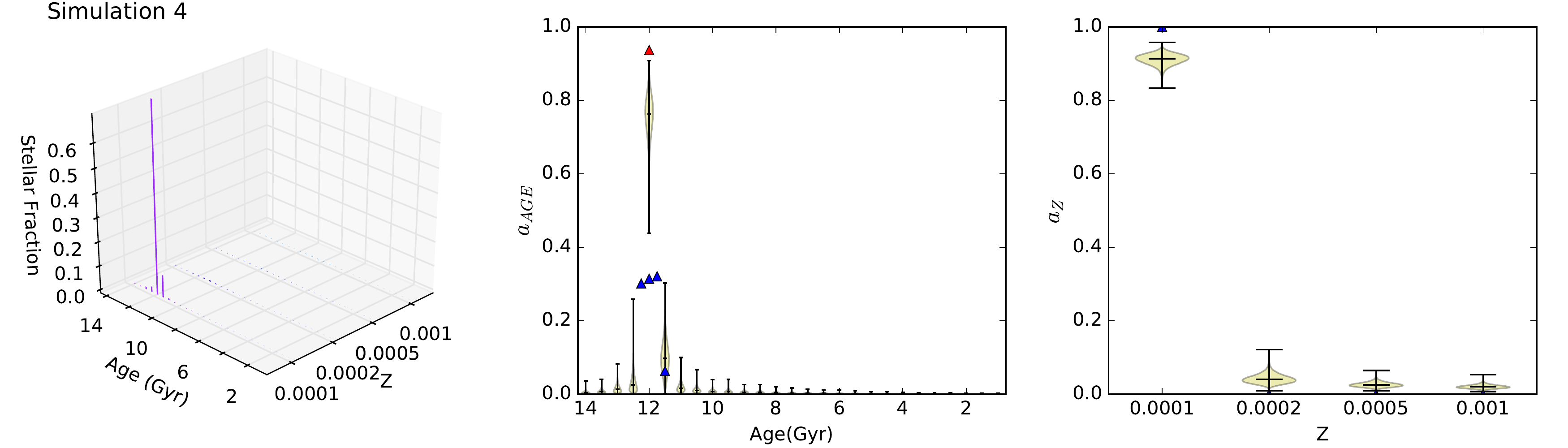}
    \includegraphics[width=\textwidth,height=0.16\textheight]{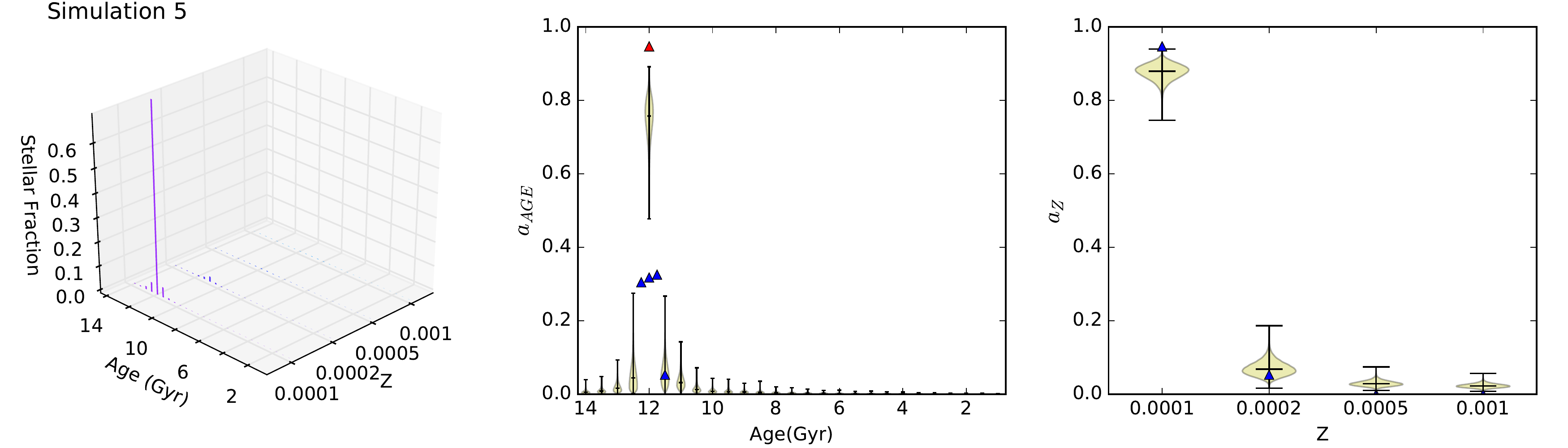}
    \caption{\label{fig:sfh_GCsimu}
    Inferred AMD for Simulations 4 ({\it upper row}) and 5 ({\it bottom row}), derived using isochrone Grid B (Table\,\ref{tab:iso}) and $\sigma^{k}_{i}\,=\,0.01$ mag.
    The {\it blue triangles} represent the input values of $a_{AGE}$ and the {\it red triangle} the sum of $a_{AGE}$ for the older population.
    }
\end{center}
\end{figure*}

\begin{figure*}
\begin{center}
    \includegraphics[width=\textwidth,height=0.16\textheight]{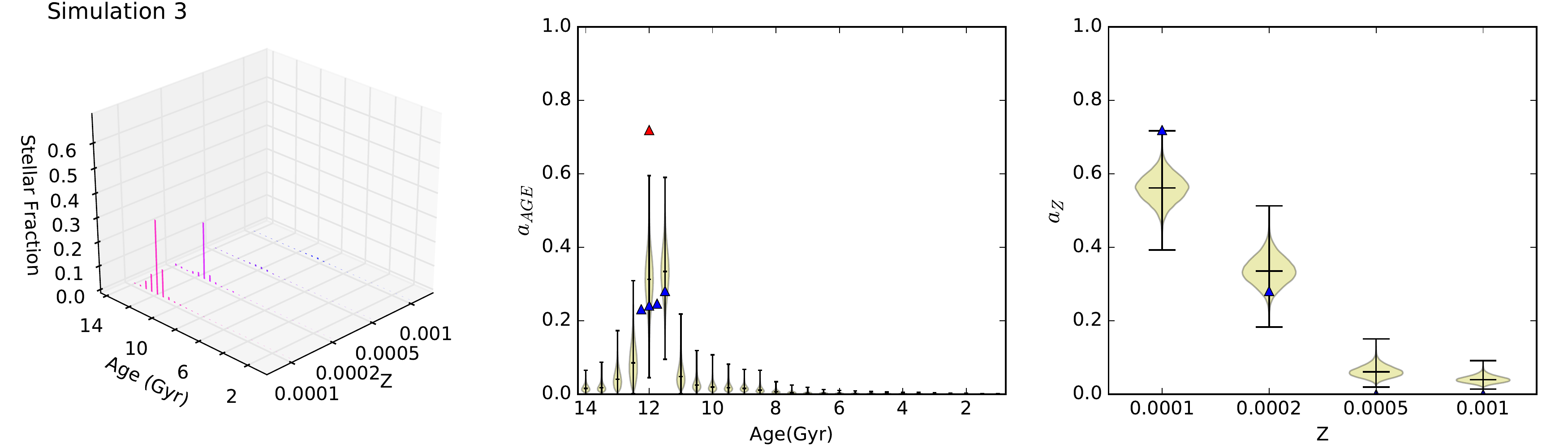}
    \includegraphics[width=\textwidth,height=0.16\textheight]{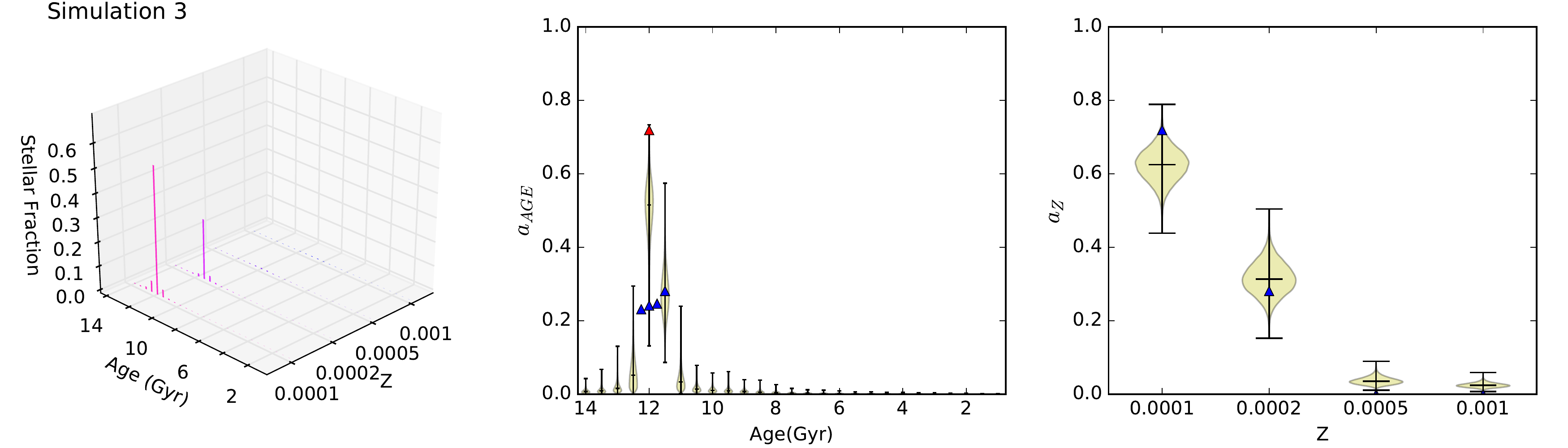}
    \caption{\label{fig:sfh_simu_2}
    Inferred AMD for Simulation 3 with photometric errors increased by 0.02 mag derived using isochrone Grid B (Table\,\ref{tab:iso}).
    ({\it Upper row}) Results for $\sigma^{k}_{i}\,=\,0.01$ mag.
    ({\it Lower row}) Results for $\sigma^{k}_{i}\,=\,0.02$ mag.
    The {\it blue triangles} represent the input values of $a_{AGE}$ and the {\it red triangle} the sum of $a_{AGE}$ for the older population.
    }
\end{center}
\end{figure*}

\subsection{Simulated CMDs}

In this section we explore the minimal separation in age, metallicity and number of stars, (i.e., {\it resolution}) that we can detect with our Bayesian inference scheme using a set of simulated stellar populations.
In these simulations we assume two star formation episodes, listed in Table\,\ref{tab:a1}:
{\it (a)} a {\it long} constant star formation rate burst starting $12.25$\,Gyr ago and ending $11.75$\,Gyr ago, with a burst duration of $0.5$\,Gyr; and
{\it (b)} an {\it instantaneous burst} happening 0.25\,Gyr or 0.75\,Gyr after the end of the previous event.

For simplicity we use the {\it simulated population} option of the PARSEC isochrone web interface\footnote{\url{http://stev.oapd.inaf.it/cgi-bin/cmd}} to generate the simulated populations.
Stars in event {\it (a)} belong to three isochrones, dated $11.75$,\,$12$ and $12.25$\,Gyr.
Stars in event {\it (b)} belong to a single isochrone, dated either $11$\,or\,$11.5$\,Gyr.
The PARSEC simulations assume the \cite{kr01} IMF.
The number of stars formed in event {\it (a)} is $\approx$\,$2.5$\,times the number of stars formed in event {\it (b)} for simulations 1,\,2\,and\,3,\,and $\approx$\,$16$\,times for
simulations 4\,and\,5 (see Table\,\ref{tab:a1}).

We obtain the apparent magnitudes $(\mathfrak{m}_{F475W},\,\mathfrak{m}_{F814w})$ of each star adding the distance modulus of Eri II, $\mu$\,=\,$22.8$ mag \citep{crnojevic:16}
to the absolute magnitudes $(M_{\rm F475W},\,M_{\rm F814W})$ output by the simulations in the Vega magnitude system.
We add the photometric error to the magnitudes of each star by random sampling a normal distribution: ${\rm F475W}'\sim\mathcal{N}(\mathfrak{m}_{F475W}\vert \delta_{F475W})$ and ${\rm F814W'}\sim\mathcal{N}(\mathfrak{m}_{F814W}\vert \delta_{F814W})$, where the prime symbols refer to the simulated apparent magnitudes. 
The magnitude errors $\delta_{\rm F475W}$ and $\delta_{\rm F814W}$ were derived from the G16 data set using the formula $\log{\delta^{k}}=c_{1} F^{k}+c_{2}$ (Fig.\,\ref{fig:a11}).
Fig.\,\ref{fig:a12} shows the resulting CMD for simulation 3 in Table\,\ref{tab:a1}.

\subsection{Recovering the AMD}

We apply the inference process described in Sections\,\ref{sec:method}\,and\,\ref{sec:age_metal} to the simulated populations using isochrone Grid B from Table \ref{tab:iso}.
Fig.\,\ref{fig:sfh_simu} shows the AMD for Simulations 1,\,2\,and\,3. For Simulation 1 the {\it left} and {\it central} panels show that the 11 Gyr population is clearly separated from the older population. 
For Simulations 2 and 3 the contribution at 11.5 Gyr is identified but the time resolution of Grid B (0.5\,Gyr) is not enough to assign the correct age to the younger population,
which looks as part of the older population in our solution for $a_{AGE}$.
However, for Simulation 3 the fact that the younger population has a different metallicity than the old population allows its clear identification in the 3D plot on the left hand
side in Fig.\,\ref{fig:sfh_simu}. For these 3 simulations we infer the true $a_Z$ distributions.
Simulations 4 and 5 are similar to Simulations 2 and 3 but the number of stars in the younger burst is reduced by more than 80\%. 
In Fig.\,\ref{fig:sfh_GCsimu} the young population is barely identified in $a_{AGE}$ and not at all in $a_Z$.

In Fig.\,\ref{fig:sfh_simu_2} we show the AMD inferred for Simulation 3 when the photometric errors are increased to $\sqrt{(\sigma^{k})^2+0.02^2}$ mag.
The most obvious effect of increasing the errors in the CMD is to broaden the MSTO, the RGB and the HB (cf. {\it right hand side} and {\it left hand side} panels
of Fig.\,\ref{fig:a12}). 
The {\it upper row} of Fig.\,\ref{fig:sfh_simu_2} shows the AMD recovered using isochrone Grid B and $\sigma^{k}_{i}\,=\,0.01$ mag, whereas the {\it lower row} corresponds to the
AMD recovered using the same isochrone grid but $\sigma^{k}_{i}\,=\,0.02$ mag.
It is clear that the solution for $\sigma^{k}_{i}\,=\,0.02$ mag is closer to the true value, especially for $a_{AGE}$.

From this experiment we conclude that we can recover correctly the age of stellar populations of the same metallicity born in different star formation episodes separated in 
time by more than the time resolution of the isochrone grid in use. 
Nearly coeval populations of different metallicity can be characterized as long as the number of stars in the less massive population is significant.
The size of the photometric errors included in the statistical model must resemble the true errors.

\bsp	
\label{lastpage}
\end{document}